\documentclass[useAMS,usenatbib]{mn2e}

\usepackage{graphicx}

\usepackage[font=small]{caption}

\title[Rotation curves in MOND]{Declining rotation curves of galaxies as a test of gravitational theory }

\author[Haghi et al.]
{Hosein Haghi$^{1}$\thanks{
E-mail:  \mbox{haghi@iasbs.ac.ir} (HH)
 },  Amir E. Bazkiaei$^{1}$, Akram Hasani Zonoozi$^{1}$, Pavel Kroupa$^{2}$\\\\
$^{1}$Department of Physics, Institute for Advanced Studies in Basic Sciences (IASBS), P.O. Box 11365-9161, Zanjan, Iran\\
$^{2}$Helmhotz-Institute f\"ur Strahlen-und Kernphysik (HISKP), Universit\"at Bonn, Nussallee 14-16, D-53115 Bonn, Germany\\}

\begin{document}

\date{Accepted \ldots. Received \ldots; in original form \ldots}

\pagerange{\pageref{firstpage}--\pageref{lastpage}} \pubyear{2013}

\maketitle

\label{firstpage} 

\maketitle

\begin{abstract}

Unlike Newtonian dynamics which is linear and obeys the strong equivalence principle, in any nonlinear gravitation such as Milgromian dynamics (MOND), the strong version of the equivalence principle is violated and the gravitational dynamics of a system is influenced by the external gravitational field in which it is embedded. This so called External Field Effect (EFE) is one of the important implications of MOND and  provides a special context to test Milgromian dynamics. Here,  we study the rotation curves (RCs) of 18 spiral galaxies and find that their shapes constrain the EFE. We show that the EFE can successfully remedy the overestimation of rotation velocities in 80\% of the sample galaxies in Milgromian dynamics fits by decreasing the velocity in the outer part of the RCs. We compare the implied external field with the gravitational field for non-negligible nearby sources of each individual galaxy and find that in many cases it is compatible with the EFE within the uncertainties. We therefore argue that in the framework of Milgromian dynamics, one can constrain the gravitational field induced from the environment of galaxies using their RCs. We finally show that taking into account the EFE yields more realistic values for the stellar mass-to-light ratio in terms of stellar population synthesis than the ones implied without the EFE.

\end{abstract}


\section{Introduction}
The observations of rotational velocities of spiral galaxies reveal significant discrepancies from Newtonian
theory such that the observed and predicted, using the
observed baryonic matter, kinematics always do not match \citep{rub70, rub78, bos78}. According to Newtonian gravity, the rotational velocity falls with distance from the center of a galaxy (the so-called Keplerian fall-off), while the observed data usually show an asymptotically flat rotation curve out to the furthest observationally accessible data points. One solution to solve this problem is assuming a dark matter halo distributed around each galaxy \citep{beg91,per96,chem11}.

Dark matter had also been inferred to contribute significantly on larger scales in the Universe from the large velocity dispersions observed in galaxy clusters by \cite{Zwic} in the 1930s, gravitational lensing of background objects by galaxy clusters such as the Bullet Cluster, the temperature distribution of hot gas in galaxies and clusters of galaxies, and more recently by the pattern of anisotropies in the cosmic microwave background.

Although the currently favored cold dark matter (CDM) model is understood to be successful on large scales (but see \citealt{kro15}), after much experimental effort up to date, no direct evidence for the existence of dark matter has been found. Moreover high resolution N-body simulations of structure formation remain in contradiction with observations and predict significantly more satellites than seen and also a wrong spatial distribution of sub-halos (e.g. \citealt{Pawlowski12, Ibata14}). \cite{kro10} have compared the predictions of the concordance cosmological model of the structures in the environment of large spiral galaxies with observed properties of Local Group galaxies and have shown that there exist prominent observational challenges for the CDM model, which might point towards the necessity of an alternative model. This inspires astronomers and physicists to search for other explanations of the discrepancy between the Newtonian dynamical mass and the luminous mass.

An alternative approach is to replace cold dark matter (CDM) by a modification of the Newtonian dynamics known as MOND \citep{mil83a,mil83b,mil83c,bek84}, (see \citealt{fam12} for a thorough review). Within this classical Milgromian dynamics framework, the Newtonian gravitational acceleration $g_N$ is replaced in the spherically symmetric case with $g=\sqrt{g_Na_0}$ when the gravitational acceleration is far smaller than the critical acceleration $a_0=1.2\times 10^{-10}ms^{-2}$ (e.g.  \citealt{mil83a,mil83b,mil83c,bek84,san02,bek06,mil08}).
Bekenstein's (2004) TeVeS theory was a mile stone in providing a consistent
relativistic foundation by introducing a tensor, a vector, and a
scalar field into the field equations. For an alternative approach to modified gravity see Moffat (2005, 2006).

Rotation curves of spiral galaxies have been studied for several
decades now, and provide a valuable body of data to determine the
radial dependency of the gravitational forces in galactic
scales. Indeed Milgromian dynamics is scale-invariant and thus yields flat rotation curves (Milgrom 2000; see also Wu \& Kroupa 2015; Kroupa 2015) and the Tully-Fisher relation \footnote{The observed
correlation between the asymptotic rotation speed and the total
baryonic mass of galaxies, $V_{\infty}^4 \propto M$, which is
not well understood in the context of dark matter.}
using only the observed distribution of visible matter and reasonable
assumptions about the stellar mass-to-light ratio as input data
\citep{beg91,san96,san02}.
When Milgromian dynamics was proposed, rotation curves of a few high surface brightness (HSB)
galaxies had been observed. Since
then, however, rotation curves for low surface brightness (LSB)
galaxies have become available which provide a more stringent and successful
test of the prediction of Milgromian dynamics than the earlier data.

Although the flattening of galactic rotation curves was
one of the first evidence of dark matter, a detailed comparison between
MOND and CDM \citep{nav96,moo99,nav04} predictions for a sample of spiral galaxies
with accurately measured rotation curves concluded that the CDM hypothesis  fails to reproduce
observed rotation curves (see, e.g., \citealt{deb01, deb02, gen04, gen05, gen07b, gen07c, mcg07, zon10, wu15}). But on the other hand, MOND tightly fits
the observations without dark matter in different types of galaxies (e.g. \citealt{mil03, fam07, gen07a, gen07b, nip07, san07, tir07, che08, wu15}) and provides the best available description. It should be emphasized that in MOND, rotation curves
are constructed with only one slightly adjustable parameter, the
stellar mass-to-light ratio, while in most CDM models, two
additional model parameters are needed to describe the dark
component.

\cite{zon10} constructed rotation curves of a large sample of
galaxies from the distribution of their detectable matter through
a set of different gravity models. While the different models reproduce the observed data
with reasonable detail, on a deeper examination, they found
significant disparities in their predictions of stellar
mass-to-light, $M_*/L$, ratios and showed that the stellar population
synthesis, SPS, analysis and the color $M_*/L$ correlation predicted
therein through various initial mass functions (IMF) could
differentiate between the models.

MOND also has been formulated as a generalization of the Poisson equation for the Newtonian gravitational field. As a result of the nonlinearity of this formulation, the strong equivalence principle is violated in MOND when considering the external field in which a system is embedded (e.g. \cite{bek84, zha06, che07, fam12}). This implies that the internal gravitational dynamics of a system embedded in an external field depends on both the internal and external gravitational field. This so-called External Field Effect (EFE) which is totally absent in Newtonian dynamics, is a decisive factor to discriminate between these two dynamics. The EFE allows high velocity stars to escape from the potential of the Milky Way \citep{wu07, fam07} and reduces the global velocity dispersion of globular clusters. In MOND, the internal gravity of a satellite galaxy of the Milky Way is determined not only by the satellite's stellar distribution (and therefore mass distribution), but also by the local strength of the Milky Way's gravitational field. For a simple introduction to the EFE see \cite{wu15} and \cite{kro15}.

Most recently, \cite{Banik16} investigated analytically the solution for a point mass immersed in a  dominating constant external gravitational field. They found that the EFE breaks the spherical symmetry of the potential of a single point-mass, such that the angle between the external field direction and the direction towards the point-mass plays an important role in the total potential. As a result, the gravity due to the mass is not always directed towards it.

Since  spiral galaxies may be influenced by the
external fields of other nearby galaxies or by the host cluster of
galaxies, generally, one can not ignore the EFE in the analysis of rotation curves. In most studies of rotation curves of spiral galaxies in MOND that
have been done up to now, the authors ignored the external field
effect. \cite{gen07a} have considered this effect for
three tidal dwarf galaxies (TDGs). They considered the EFE to
show that the MOND framework is very likely to correctly explain
the kinematics of the TDGs. \cite{wu08} calculated the rotation curve of the MW within the MOND framework, allowing for
various values of the external field.  Recently,  \cite{wu15} applied the external field effect to predict the masses of galaxies in the proximity of the dwarf galaxies compiled by \cite{ler014} assuming they are embedded in an external field of a companion galaxy or a cluster of galaxies.
Most recently, \cite{hee15} presented an analysis of the galactic rotation curves taking into account the EFE and showed that the EFE can significantly improve some galactic rotation curve fits by decreasing the predicted
velocities of the external part of the rotation curves.

Although MOND successfully reproduces the observed rotation curves of a large sample of galaxies, there are some galaxies with declining rotation curves in the outer parts not favorable to MOND. The aim of the present work is to investigate the rotation curves of a sample of galaxies with a wide range of luminosities and morphologies from  dwarf irregulars to bright spirals under the MONDian EFE.

The paper is organized as follow: the basics of MOND and the EFE are discussed in Sections \ref{basic} and \ref{efe}, respectively.  We describe the
rotation curve data of a sample of galaxies in Section \ref{data}. In Section \ref{fit} we produce the rotation curve fits to the used sample of galaxies under the influence of the EFE. This is followed by a presentation of the results of fits in Section \ref{result}. Conclusions are presented in Section \ref{conclusion}.

\section{The MOND basics}\label{basic}

In the non-relativistic version of MOND, the acceleration of an isolated system, $g$ is related to
the Newtonian acceleration $g_N$ through the following relation\footnote{Note that Eq. 1 is valid in spherically and axially symmetric cases.}:

\begin{equation}
\textbf{g} \mu(\frac{g}{a_0})= \textbf{g}_N, \label{Mond}
\end{equation}
where  $\mu(x)$ is an interpolation function which runs smoothly
from $\mu(x)=x$ at $x\ll1$ to $\mu(x)=1$ at $x\gg1$. The standard
interpolating function is $\mu_1(x)=x/(\sqrt{1+x^2})$, but
\cite{fam05} suggested the simple function
$\mu_2(x)=x/(1+x)$, which provides a better fit to the
rotation curve of the Milky Way. The simple function is
compatible with the relativistic theory of MOND (TeVeS) put
forward by \cite{bek04}. The typical value for the parameter
$a_0$, obtained from an analysis of a sample of spiral galaxies with
high-quality rotation curves, is
$a_{0}=1.2\pm0.27\times10^{-10}ms^{-2}$ \citep{beg91}.
\cite{fam07} showed that using the simple function leads to a
slightly different value of $a_{0}=1.35\times10^{-10}ms^{-2}$.  The first value corresponds to $a_{0}=3600\pm810~pcMyr^{-2}$ and the second value to $a_{0}=4050 ~pcMyr^{-2}$.

For the standard interpolating function,  one can obtain the MOND acceleration, ${\bf g}$
in terms of  the Newtonian acceleration ${\bf g_N}$ as follow:

\begin{equation}
{\bf g}={\bf g}_N\sqrt{ \frac{1}{2}+\frac{1}{2}\sqrt{1+4(\frac{a_0}{g_N})^2}}.
\label{amond2}
\end{equation}
Here, $g_N=\frac{MG}{r}$, $r$ is the radius and  $M=M_{d}+M_{b}+M_{g}$, includes the total  stellar disk, bulge, and gaseous disk, respectively. The amplitude of $M_{d}$ and $M_{b}$, which are determined by photometric observations, can be scaled according to the chosen, or fitted, stellar mass-to-light ($M_*/L$) ratio. The value of $M_{g}$ is derived from HI observations, when they are available. Therefore, rotation curves in MOND can be expressed as:

 \begin{equation}
v_{MOND}^2=v^2_N\sqrt{ \frac{1}{2}+\frac{1}{2}\sqrt{1+4(\frac{a_0}{g_N})^2}}.
\label{v2}
\end{equation}

\section{EFE in MOND}\label{efe}

In the context of MOND,  the internal dynamics of a stellar system is
affected both by the internal and external fields, due to the violation of the strong
equivalence principle \citep{bek84}. In classical
Newton/Einstein gravity, the situation is entirely different,
i.e., a uniform external field does not affect the internal
dynamics of a stellar system. If the external field becomes comparable to or larger than $a_0$, a galaxy embedded in this external field will always be in the Newtonian regime, even if the internal accelerations are low. In
practice, no objects are truly isolated in the Universe and they
may be influenced by other systems (e.g., from a nearby galaxy, or large scale structure). As a consequence of the EFE, the predicted velocities of the outer part of the rotation curves of galaxies where the internal acceleration becomes equal to the external acceleration, decreases \citep{gen07a,wu08}. This is equivalent to the truncation of the phantom dark matter halo \citep{wu15}.

For a galaxy with a density distribution $\rho_{g}$ which is
embedded in a host galaxy cluster (or located nearby another galaxy) with a density distribution
$\rho_{ext}$, the acceleration of stars in the galaxy satisfies
the generalized Poisson equation

\begin{equation}
 \nabla.[\mu(\frac{\nabla\Phi}{a_{0}})\nabla\Phi] = 4\pi G \rho_t = \nabla^2\Phi_N , \label{mon1}
\end{equation}
where $\nabla\Phi$ is the Milgromian potential generated by the total
matter density, $\rho_{t}=\rho_g+\rho_{ext}$ and $\Phi_N$ is the Newtonian gravitational potential solution of the standard Poisson equation.
One way to solve Eq. \ref{mon1} is
then to assume that the total acceleration is the sum of the
internal $a_i$ and the external $a_e$ acceleration which both
satisfy the generalized Poisson equation as
\citep{bek84,wu07,wu08,ang08},

\begin{equation}
 \nabla.[\mu(\frac{|\bf{a_e}+\bf{a_i}|}{a_{0}})({\bf a_e}+{\bf a_i})] = 4\pi G \rho_g. \label{mon2}
\end{equation}
In Eq. \ref{mon2}, the direction of the external acceleration is important
for the dynamics of stars in the clusters. For example, for two
stars the acceleration would be different if $a_e$ is parallel or
orthogonal to their internal acceleration.

Several attempts to solve the generalized Milgromian Poisson equation considering the EFE have been done. \cite{wu07} simulated an isolated object with a static Milgromian
potential solver, but changed the boundary condition on the
outermost grid point to be nonzero. \cite{fam07}
estimated the escape velocity of galaxies in Milgromian dynamics and assumed
$\mu(|\bf{a_e}+\bf{a_i}|/a_{0}){\bf a_i}={\bf a_N}$. As a first
order test they replaced $|\bf{a_e}+\bf{a_i}|$ with $(a_e+a_i)$,
or $\sqrt{a_i^2+a_e^2}$. This is an approximation, which reduces a 3-dimensional problem to a one-dimensional one, neglecting the general relative orientations of the acceleration vectors (i.e. a possible angular difference between $\bf{a_i}$ and $\bf{a_e}$).

Using the Milgromian dynamics N-body code, NMODY \citep{Londrillo09},  \cite{hag09} calculated the evolution of globular cluster-scale stellar systems under the influence of the external field of a host galaxy by adding the constant external gravitational field to the internal acceleration inside the $\mu$-function, considering the different angle between external and internal acceleration for all stars throughout the evolution. This method was the first attempt to include the EFE  in the Milgromian dynamics N-body code to investigate the internal evolution of non-isolated stellar systems.
The implied phase-transition of a star cluster moving away from its host galaxy has been studied by \cite{wu13}.

\section{The data}\label{data}

Our sample includes a collection of 18 galaxies, taken from \cite{san96}, \cite{mcg98}, \cite{san98}, and \cite{beg91}, spans a wide range of luminocities and morphological types from faint gas-rich dwarfs to bright spiral galaxies.  They are listed in Table 1 and shown in Fig. 2.  The sample includes several gas dominated low-surface brightness (LSB) galaxies, e.g. DDO 154 and IC 2574 that show large discrepancies between visible mass and their dynamical mass which makes them ideal objects to test MOND.  There are also a number of high surface brightness (HSB) galaxies with a massive stellar component and a low gas content with well-extended rotation curves, e.g.,  NGC 4100, and NGC 5033. Contrary to the LSBs with slowly rising rotation curves, the rotation curves of HSBs rise steeply to a maximum and decline slowly into an almost horizontal asymptote.
Nine galaxies of the sample  are part of THINGS that are observed in HI with the Very Large Array in the B, C and D configurations \citep{wal08}, and  their rotation curves  are derived by \cite{deb08}. Nine other members of the sample are taken from the galaxies used in the MOND fits of \cite{zon10}. Seven members of the sample are LSB and 11 of them are HSB. Three of the galaxies listed in Table 1 have a central bulge and are treated differently, to be explained shortly.

\begin{figure}
\includegraphics[width=85mm]{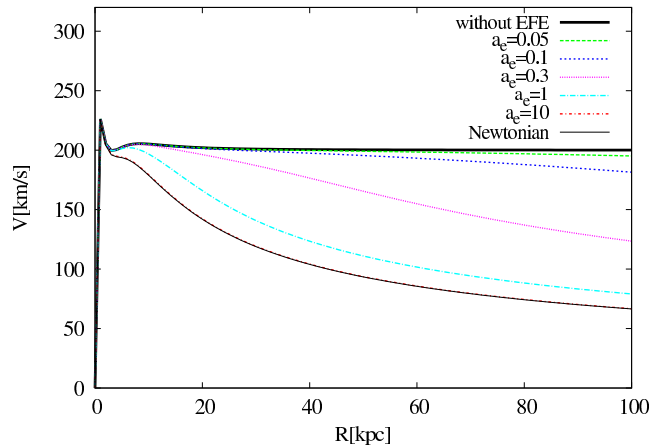}
\caption{The calculated rotational velocity curve for a typical disk galaxy containing a stellar bulge and disk with the total mass of $M=10^{11} M_{\odot}$, in Newtonian and MOND dynamics and in different external gravitational fields ($a_e$) in units of $a_0$.  Rotation curves of a galaxy embedded in a stronger external field ($a_e>a_0$) show the return to a Keplerian behavior at large radii. The case of $a_e=10$ is completely Newtonian such that the corresponding curve is hidden behind the Newtonian curve. See Sec. 6.1 for further details.
} \label{f1}
\end{figure}

\section{Fitting the Rotation Curves}\label{fit}

To calculate the MOND rotation curves we used the same method as used in \cite{san98}. However, in principle, the Milgromian dynamics Poisson equation should be used to calculate the Milgromian dynamics circular speed, but \cite{mil86} has shown that the results derived for the field equation slightly differ ($\leq 5\%$) from those using the original Milgromian dynamics prescription.  The first step is to determine the Newtonian acceleration of the detectable matter,
$g_N$ via the classical Poisson equation. The stellar and gaseous mass distribution are assumed to be in a thin disk. We corrected for the mass fraction of helium by scaling the HI mass by a factor of 1.4. We assume  the HI gas is in co-planer rotation about the center of the galaxy, an assumption
which may not hold in galaxies with strong bars \citep{san02}.

Given the Newtonian acceleration, the effective acceleration without EFE and corresponding rotational velocity is calculated from the MOND formulae, i.e., Eqs. \ref{amond2}  and \ref{v2}, respectively. We assume that the stellar mass-to-light ratio ($M_*/L$) is constant with radius, throughout the galaxy, though
this is not strictly the case, because of the color gradient in
spiral galaxies.  The integrated galactic initial mass function (IGIMF) theory \citep{weid13, kro13, Pfl08} provides a computable quantification of the radial variation of the IMF in a disk galaxy, and age and metallicity gradients also imply  a radially changing $M^*/L$ value. This is to be studied in the future.

Using the standard interpolation function and considering the EFE, a little algebra shows that the circular speed for MOND can be expressed as:

\begin{equation}
v_{MOND}^2=v_{N}^2\sqrt{   \frac{   1-(\frac{a_e}{g_N})^2 + \sqrt{(1-(\frac{a_e}{g_N})^2)^2+ \frac{4(a_0^2+a_e^2)}{g_N^2}}  } {2}  },
\end{equation}
\label{efe-fit}

where,  $v_{N}^2=v_{d}^2(\frac{M^*}{L})_d + v_{b}^2(\frac{M^*}{L})_b + v_{g}^2$, with $v_d$,  $v_b$, and $v_g$ the Newtonian contribution of the stellar disk, bulge and gas to the rotation curve, respectively. $(\frac{M^*}{L})_d$ and $(\frac{M^*}{L})_b$ are the stellar mass-to-light ratios of the disk and the bulge.
How  $v_{MOND}$ depends on the EFE is demonstrated in Fig. 1 (see Sec. 6.1).

As an approximation  to the MOND rotation curve, we used the 1-D version of Eq. \ref{mon2} which has been shown to be a good approximation of the true 3-D version \cite{fam07, wu07} and we assume that $a_e$ is perpendicular to the disk-plane of the galaxy.

To fit the observational velocity curve with the theoretical model we employ the $\chi^2$ goodness-of-fit test. Fitting of the calculated rotation curves to the observed data points is
achieved by adjusting the $M_*/L$ ratio and the external gravitational field (in the case of a fit with the EFE), by minimizing the reduced least-squares value
\begin{equation}
 \chi^2=\frac{1}{(N-P-1)}\sum_{i=1}^{N}\frac{(v^{i}_{Mond}-v_{obs}^{i})^2}{\sigma_i^2}\label{chi},
\end{equation}
\noindent where $\sigma_i$ is the observational uncertainty in the rotation speeds and $P$ is the number of degrees of freedom. $N$ is the number of observed velocity values along the radial direction in a galaxy. The $M_*/L$ ratio of the stellar component and $a_e$ are free parameters (i.e., P=2), where constrained values must however be physically relevant.

\section{Results} \label{result}

Here, in this section we present the results of Milgromian dynamics fits. Before addressing the Milgromian dynamics rotation curves, first we assess the general effect of the external field on the shape of a rotation curve in Milgromian dynamics.

\subsection{Rotation curves with the EFE: The General Case}

An external gravitational field on a galaxy can be induced by
nearby galaxies or clusters of galaxies in which the galaxy is embedded, and consequently it can cause a suppression of
the boost to gravity due to MOND such that the rotation curve become Newtonian. Particularly, low surface brightness galaxies (with internal acceleration much smaller than $a_0$) are more affected than others. (c.f. \citealt{wu15}).

In order to demonstrate the EFE  on the shape of a typical rotation curve, for different values of the external gravitational field (in the range $a_{e}=$0.1 to 10 $a_0$), we calculate the circular velocity of a galaxy using Eq. 6 that consists of two idealized stellar components including a disc potential given by
\begin{equation}
\Phi_d (x,y,z) = -\frac{G M_d}{\sqrt{x^2+y^2+\left(a + \sqrt{z^2+b^2}\right)^2}},
\end{equation}
and a central bulge,  given by
\begin{equation}
	\Phi_b (x,y,z) = -\frac{G M_b}{r+r_c},
\end{equation}
where $M_d = 7.5\times10^{10} M_{\odot}$, $a=5.4$ kpc,  $b=0.3$ kpc, $M_b = 2.5\times10^{10} M_{\odot}$, and $r_c=0.5$ kpc.

There is a remarkable difference in predicted rotational velocity in MOND with different values of $a_e$ as shown in Figure \ref{f1}. The EFE has a significant effect on the rotational velocities at the outskirts of galaxies.  In all rotation curves the Milgromian dynamics velocity curves lie above the Newtonian one. As the external field gets stronger the rotation curve shifts to the Newtonian prediction.  In particular, the external field erases the flat behavior of the rotational velocity in the outer parts.  As will be discussed with more detail in the next section, the EFE leads to a higher mass-to-light ratio in Milgromian dynamics because of the weakening of the effect of Milgromian dynamics by pushing the rotation curve towards the Newtonian curve, especially in the outer parts. An unphysical fit to a given falling rotation curve would thus require an unphysically large  $M_*/L$ ratio. This allows us to assess if fits  which are formally good are also physically acceptable.

\subsection{MOND fits with and without the EFE}

In this section we compare the calculated circular velocity curves in the
framework of Milgromian dynamics in two different regimes (with and without the EFE) with the
observed rotation curves. First we perform Milgromian dynamics fits neglecting the EFE by
fitting Eq. \ref{v2} to the observed rotation curves with a least square
algorithm such that P=1 in Eq. 7. The Milgromian dynamics acceleration is fixed at its standard value ($a_0=1.2
\times 10^{-10} ms^{-2}$ ) and Milgromian dynamics fits are made with only the $M_*/L$ ratio
as a single free parameter in this case.

In  Fig. \ref{f2} we show the results  of the one-parameter MOND fits of
theoretically constructed RCs to the observations of 18 galaxies. In each
panel, the measured rotational velocities are indicated as red points with
error-bars and the best-fitting Milgromian dynamics RCs without the EFE are shown
by blue-dotted lines. The stellar disk and gas contributions are
represented as the pink dotted-line and green dashed line, respectively.  In
Tables \ref{t1} and \ref{t2}, we summarize the key results of the fitted
RCs, that
is, $\chi^2$, and the implied $M_*/L$ ratios of the stellar components.

Three galaxies have a prominent bulge component. One expects a bulge with
an older population of stars to have a higher
$M_*/L$ ratio than a disk with a younger population. Therefore, to obtain
a more realistic representation for these galaxies, we have allowed the model
to choose different $M_*/L$s for the disk and the bulge. The result, shown
in Table \ref{t1}, confirms the expectation.  For NGC 4736 and 5055, $(M_*/L)_d<(M_*/L)_b$, suggesting the bulge to be older than the disk, as is observed for the MW. However, for NGC 3031 $(M_*/L)_d>(M_*/L)_b$ which may arise if its bulge is younger that the disk suggesting a possibility to test this prediction by further
observation.

Discrepancies between the RCs predicted by Milgromian dynamics without the EFE and the
observed one are seen for most of the galaxies in the sample.  In
particular, Milgromian dynamics without the EFE provides acceptable fits ($\chi^2\leq 2$) for
only 5 out of the 18 galaxies in our sample (i.e., NGC 2366, NGC 3621, NGC
3769, NGC 4183, UGC 6983), and for the other 13 galaxies, MOND
overestimates the rotation velocities of larger $R$.

The obtained best-fitting Milgromian dynamics RCs including the EFE are also indicated by the
thick lines.
Figure \ref{f2} shows that there is an observable difference
between Milgromian dynamics in RCs with and without the EFE.  In all RCs the velocity curve
of MOND with the EFE is above the Milgromian dynamics one without the EFE in the central part
of the galaxies, but falls underneath in the outer parts (see e.g., UGC 6446 and
UGC 6983 in Fig. \ref{f2}). Adding the EFE improves the
RCs fits in all galaxies.

The EFE improves the rotation curve fits of 14 out of the 18 galaxies in
the sample (i.e., achieves an acceptable fit with $\chi^2\leq 2$). In
roughly half of the galaxies (DDO 154, IC 2574, NGC 3198, NGC 5533, NGC
5033, NGC 2998,  NGC 4183, UGC 6983, and UGC 6446) for which the quality of the
fit was quiet poor when neglecting the EFE, the rotation curves fit
significantly improved such that the corresponding values of $\chi^2$
decrease by a factor of 2 or 3. In particular,  for the cases that the
Milgromian dynamics (without EFE) curve falls bellow the observed rotation curve in the
inner part, and above the rotation curve in the outer region the EFE helps to improve the quality of the Milgromian dynamics fit to the observation.  The case of
the galaxy UGC 6446 (Fig. \ref{f2}) is a very interesting example. As can be
seen, the fit is improved at both small radii because of an increase of the
inner part of the rotation curve owing to the increasing of the $M_*/L$ ratio
in Milgromian dynamics with the EFE and also in the outer part where the rotational
speed decreases due to the EFE.

The implied $M_*/L$ ratios are given in Tables 1 and 2.  Accounting for $a_e\neq0$
leads to a higher estimation of $M_*/L$. The uncertainties on the
best-fitted values of $M_*/L$ and $a_e$ have been derived from the 68\%
confidence level.  The marginalized 68\%  confidence intervals on the
parameters $a_e$ and $M_*/L$ are indicated in the central and right panels of
Fig. \ref{f2} by a horizontal line.  The 1 and 2 $\sigma$ confidence regions for the best-fit parameters to the observed rotation curves of 18 dwarf and spiral galaxies listed in Tables 1 and 2 are shown in Fig. \ref{sigma}. The central points correspond to the best-fit values.

In Figure \ref{f6}, we compare the fitted global disk $M_*/L$ ratios to
the predictions of population synthesis by \cite{bel01} and \cite{por04}.
The red and green symbols are the implied $M_*/L$ ratios from the Milgromian dynamics fits
with and without the EFE, respectively.
Our results are well-compatible with the prediction of SPS in both the B- and
the K-band. The overall trend is similar - the redder the galaxy, the higher
the  $M_*/L$, but the values generally scatter around the models. We see
that the $M_*/L$ values lie within the range set by a Salpeter and a Kroupa
IMF for stars with masses between 0.1 and 100 $M_{\odot}$. According to Table 2, and the color-$M_*/L$ relation in the K-band,
i.e., $log(M_*/L_k)= 1.46(J-K)-1.38$ \citep{Oh08}, interestingly there
are three cases (DDO 154, IC 2574, and NGC 2366) for which the implied
$M_*/L$ ratio from the Milgromian dynamics fit without the EFE is dramatically lower than the
prediction of the SPS models. Including the EFE, the $M_*/L$ ratios from the
MOND fits for these three cases are in the reasonable range and in agreement
with the prediction of SPS models. 
A comparison of the best-fit stellar mass-to-light ratios obtained from MOND rotation curve fits including the EFE,
$(M_*/L)_{EFE}$, with the independent expectations of stellar population synthesis models, $(M_*/L)_{SPS}$, is shown in Fig. \ref{sps-efe}. The black line shows the region where $(M_*/L)_{SPS}=(M_*/L)_{EFE}$. A remarkably good agreement can be seen between the implied mass-to-light ratios from the MOND fits and those predicted by SPS models.

The radial distribution of $f_{gas}=M_{gas}/M_{baryon}$ in our sample of galaxies is plotted in Fig. \ref{fgas}. As can be seen, the sample includes several gas-poor ($f_{gas}\simeq0.1$) galaxies that typically have their well-extended rotation curve rise steeply to a maximum which declines
slowly into an almost horizontal asymptote. There are also a number of dwarf, gas-dominated galaxies ($f_{gas}\simeq1$).

It should be noted that there are small variations in the inclination of the galaxies \citep{deb08},  we therefore neglected the uncertainty in
the inclination of the galaxies as they are included within the errors associated with each data point we used here. As will be discussed for some galaxies in our sample in Sec. 6.3, the inclusion of uncertainties in the distance of galaxies can slightly alter the results so that a larger distance leads to a higher best fitting value of the external acceleration.

 \begin{table*}
\begin{center}
\resizebox{\textwidth}{!}{
\begin{tabular}{ccccccccccccccc}
\hline

$Galaxy$  & $J-K$ & $L_V$ & $[M^*/L_K]_{d} $&$[M^*/L_K]_{b} $&$\chi^{2}$ & $[M^*/L_K]_{d} $&$[M^*/L_K]_{b} $&$\chi^{2}$& $a_e$ & $a_{e,imposed}$\\
        &  &$[L_{\odot}]$   &$NO~EFE$   & $NO~EFE$ & $NO~EFE$     &$EFE$     & $EFE$ &$EFE$       &$[a_0]$ &$[a_0]$\\

\hline
\\
$(1)~DDO~154$ & & $4.4\times 10^7$ & $0.01^{+0.25}_{-0.01}$ && $3.10$ & $0.55^{+0.44}_{-0.36}$& & $0.35$ & $0.145^{+0.035}_{-0.030}$ & $2.87^{+3.29}_{-2.16}\times 10^{-3}$ \\
\\
$(2)~~IC~2574~$  & $0.76$ & $8.5\times 10^8$ & $0.02^{+0.08}_{-0.02}$ & & $3.86$ & $0.34^{+0.15}_{-0.13}$ && $1.13$ & $0.330^{+0.090}_{-0.075}$ & $3.43^{+2.86}_{-3.18}\times 10^{-2}$ \\
\\
$(3)NGC~2366$ & $0.67$ & $4.7\times 10^8$ & $0.04^{+0.26}_{-0.04}$ && $0.54$ & $0.33^{+0.31}_{-0.22}$ && $0.38$ & $0.240^{+0.125}_{-0.155}$ & $0.088^{+108}_{-0.068}\times 10^{-3}$ \\
\\
$(4)NGC~3031$ & $0.93$ & $2.0\times 10^{10}$ & $0.90^{+0.05}_{-0.05}$ & $0.57^{+0.23}_{-0.22}$ & $4.29$ & $1.02^{+0.05}_{-0.05}$ & $0.19^{+0.22}_{-0.19}$ & $3.30$ & $1.600^{+2.300}_{-0.965}$ & $2.14^{+30.7}_{-1.98}\times 10^{-2}$ \\
\\
$(5)NGC~3198$ & $0.95$ & $1.3\times 10^{10}$ & $0.53^{+0.08}_{-0.06}$ && $5.49$ & $0.69^{+0.08}_{-0.07}$ && $1.30$ & $0.240^{+0.050}_{-0.050}$ & $8.14^{+3950}_{-7.62}\times 10^{-3}$ \\
\\
$(6)NGC~3521$ & $0.96$& $4.4\times 10^{10}$ & $0.53^{+0.07}_{-0.06}$ && $4.90$ & $0.55^{+0.06}_{-0.06}$ && $4.12$ & $0.645^{+0.760}_{-0.645}$ & $3.37^{+14.8}_{-3.33}\times 10^{-3}$ \\
\\
$(7)NGC~3621$ & $0.88$& $1.3\times 10^{10}$  & $0.45^{+0.05}_{-0.06}$ && $1.70$ & $0.46^{+0.06}_{-0.05}$ && $1.49$ & $0.125^{+0.090}_{-0.125}$ & $5.90^{+33.9}_{-2.69}\times 10^{-4}$\\
\\
$(8)NGC~4736$ & $0.92$ & $9.8\times 10^{9}$ & $0.37^{+0.07}_{-0.07}$ & $1.15^{+0.18}_{-0.16}$ & $5.96$ & $0.55^{+0.07}_{-0.06}$ & $1.01^{+0.17}_{-0.17}$ & $1.57$ & $1.450^{+1.350}_{-0.505}$ & $3.81^{+5.86}_{-3.20}\times 10^{-3}$ \\
\\
$(9)NGC~5055$ & $0.96$& $1.8\times 10^{10}$  & $0.41^{+0.05}_{-0.05}$ & $6.23^{+1.71}_{-1.60}$ & $3.06$ & $0.46^{+0.05}_{-0.05}$ & $5.58^{+1.69}_{-1.55}$ & $1.15$ & $0.310^{+0.115}_{-0.135}$ & $1.2^{+1.02}_{-1.06}\times 10^{-2}$ \\
      \hline
\end{tabular}}
\end{center}
\caption{ The results for the Milgromian dynamics fit.  Columns~1 and 2 give the galaxy name and J-K color. Columns~3 gives the luminosity of galaxy in the V-band. Columns 4 and 5  are the best-fitting stellar mass-to-light ratios in the K-band for discs and bulges (if present) separately obtained for the Milgromian dynamics without EFE. The corresponding reduced chi-squared is presented in column 6. Best-fitted parameters for Milgromian dynamics when the external field effect is taken into account are given in columns 7, 8, and 9. Column~10 gives the optimal value for the external field. The error bars are the 68\% confidence level.  The last column gives the imposed gravitational field by non-negligible nearby sources of each individual galaxy.  \label{t1}}
\end{table*}

\begin{table*}
\begin{center}
\resizebox{\textwidth}{!}{
\begin{tabular}{ccccccccccccccc}
\hline

$Galaxy$  & $B-V$ & $L_V$ & $M^*/L_B $&$\chi^{2}$ & $M^*/L_B$ & $\chi^{2}$ & $a_e$ & $a_{e,imposed}$\\
        &   & $[L_{\odot}]$  & $NO~EFE$   &$NO~EFE$     & $EFE$   & $EFE$     & $[a_0]$ & $[a_0]$\\
\hline
\\
$(10)NGC~2998$ & $0.45$ & $2.7\times 10^{10}$ & $1.24^{+0.09}_{-0.09}$ & $2.66$ & $1.38^{+0.08}_{-0.08}$ & $1.22$ & $0.220^{+0.050}_{-0.065}$ & $1.06^{+81}_{-0.76}\times 10^{-3}$\\
\\
$(11)NGC~3769$  & $0.45$ & $4.7\times 10^9$ & $1.23^{+0.28}_{-0.25}$ & $0.80$ & $1.26^{+0.26}_{-0.24}$ & $0.81$ & $0.085^{+0.105}_{-0.085}$ & $1.53^{+470}_{-1.32}\times 10^{-3}$\\
\\
$(12)NGC~4100$  & $0.63$ & $2.1\times 10^{10}$ & $2.43^{+0.23}_{-0.23}$ & $2.15$ & $2.52^{+0.22}_{-0.22}$ & $1.87$ & $0.335^{+0.220}_{-0.335}$ & $1.08^{+2850}_{-0.72}\times 10^{-4}$\\
\\
$(13)NGC~4183$ & $0.39$ & $4.5\times 10^9$ & $0.73^{+0.20}_{-0.18}$ & $1.01$ & $0.95^{+0.19}_{-0.18}$ & $0.27$ & $0.265^{+0.110}_{-0.130}$ & $3.29^{+93.1}_{-2.86}\times 10^{-3}$\\
\\
$(14)NGC~5033$ & & $3.9\times 10^{10}$ & $4.68^{+0.21}_{-0.21}$ & $7.04$ & $4.99^{+0.20}_{-0.19}$ & $3.10$ & $0.335^{+0.070}_{-0.080}$  & $9.72^{+33.9}_{-9.57}\times 10^{-2}$\\
\\
$(15)NGC~5371$ & $0.65$ & $3.9\times 10^9$ & $1.62^{+0.08}_{-0.08}$ & $10.49$ & $1.72^{+0.07}_{-0.07}$ & $7.26$ & $0.330^{+0.080}_{-0.085}$ & $3.17^{+102}_{-2.25}\times 10^{-3}$\\
\\
$(16)NGC~5533$ & $0.77$ & $6.0\times 10^{10}$ & $3.35^{+0.34}_{-0.33}$ & $2.51$ & $3.88^{+0.32}_{-0.31}$ & $1.09$ & $0.200^{+0.050}_{-0.060}$ & $1.25^{+4550}_{-1.07}\times 10^{-3}$\\
\\
$(17)UGC~6446$ & $0.39$ & $8.7\times 10^8$ & $0.52^{+0.21}_{-0.17}$ & $2.43$ & $0.94^{+0.22}_{-0.19}$ & $0.75$ & $0.325^{+0.090}_{-0.090}$ & $5.58^{+2.31}_{-5.58}\times 10^{-2}$\\
\\
$(18)UGC~6983$ & $0.45$ & $2.4\times 10^9$ & $1.74^{+0.40}_{-0.35}$ & $1.37$ & $2.25^{+0.37}_{-0.34}$ & $0.79$ & $0.390^{+0.155}_{-0.175}$ & $5.58^{+2.31}_{-5.58}\times 10^{-2}$\\
\\
      \hline
\end{tabular}}
\end{center}
\caption{ The same as Table 1, but the luminosities and colors are measured in the V-band and in B-V, respectively. \label{t2}}
\end{table*}

\subsection{Remarks on the environment of individual galaxies}

\begin{figure*}
\includegraphics[width=180mm]{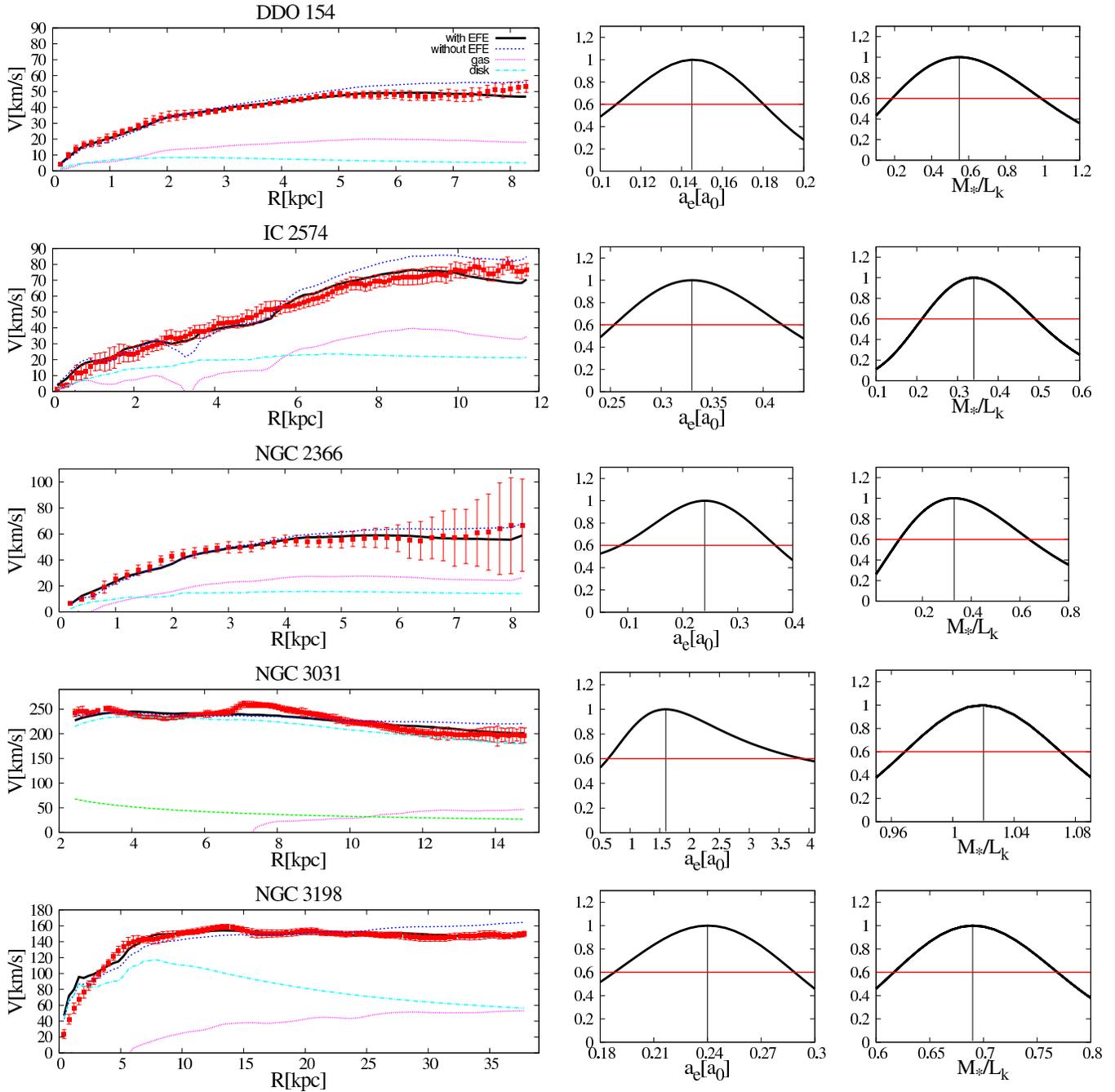}
\caption{ Milgromian dynamics fits to the observed rotation curves of 18 dwarf and spiral galaxies listed in Tables 1 and 2. In the left panels data points represent the measured rotational velocities and their errors. The pink dotted line shows the contribution of the HI to the rotation curve. The dashed and the green dotted curves give, respectively,  the contribution of the stellar disc and bulge in rotational velocity to the best-fitting $M_*/L$ values. The blue dotted line is the best MOND fit without EFE and with only the $M_*/L$ ratio as a free parameter. The solid black lines give the MOND best fit with both $M_*/L$ ratio and external field, $a_e$, as free parameters. The details of best-fitting parameters are listed in Tables 1 and 2. The middle and right panels indicate the marginalized 68\% confidence levels on the free parameters, $a_e$ and $M_*/L$ ratio, respectively for the MOND fit with EFE.
} \label{f2}
\end{figure*}

\begin{figure*}
\ContinuedFloat
\includegraphics[width=180mm]{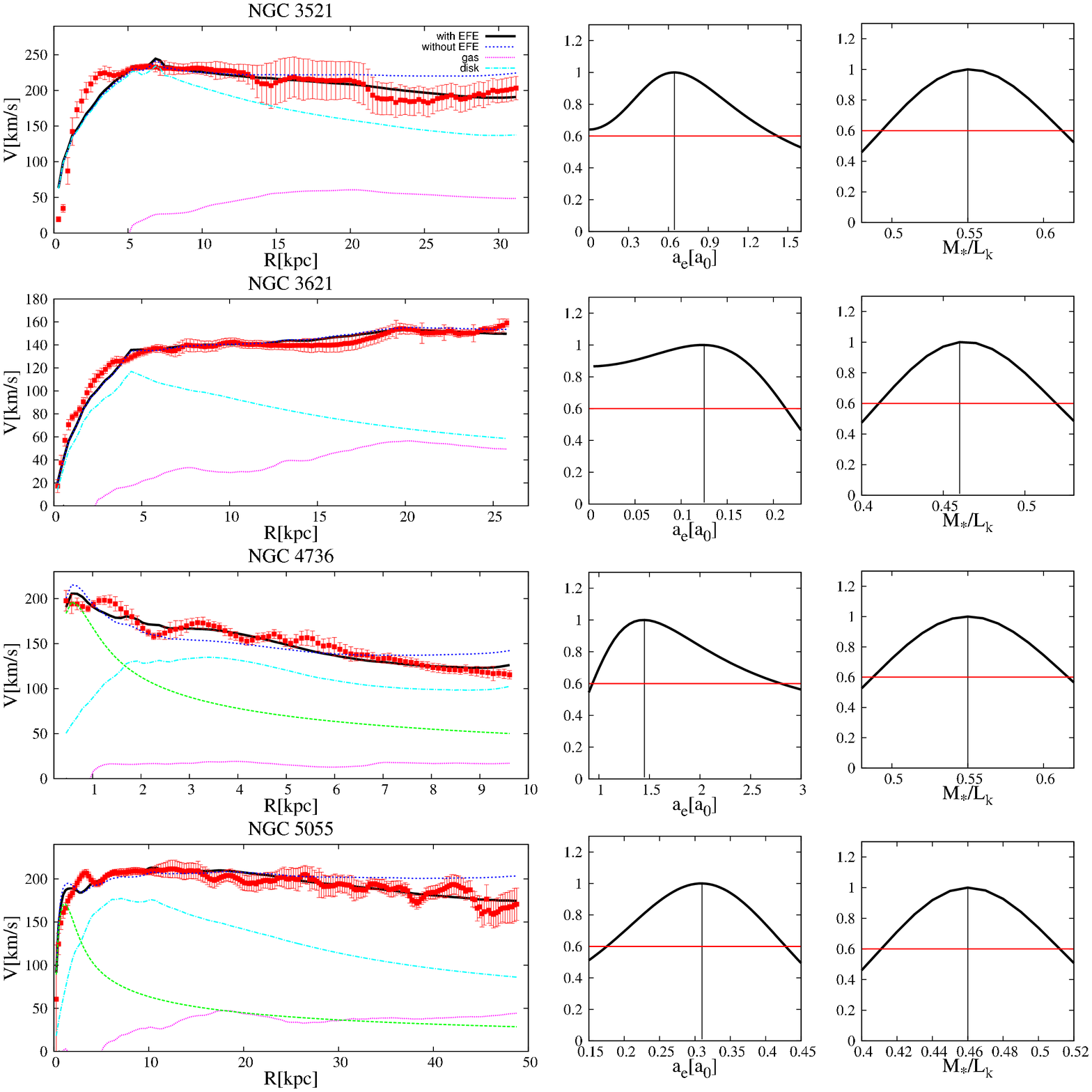}
\caption{Continued.
} \label{f3}
\end{figure*}

\begin{figure*}
\ContinuedFloat
\includegraphics[width=180mm]{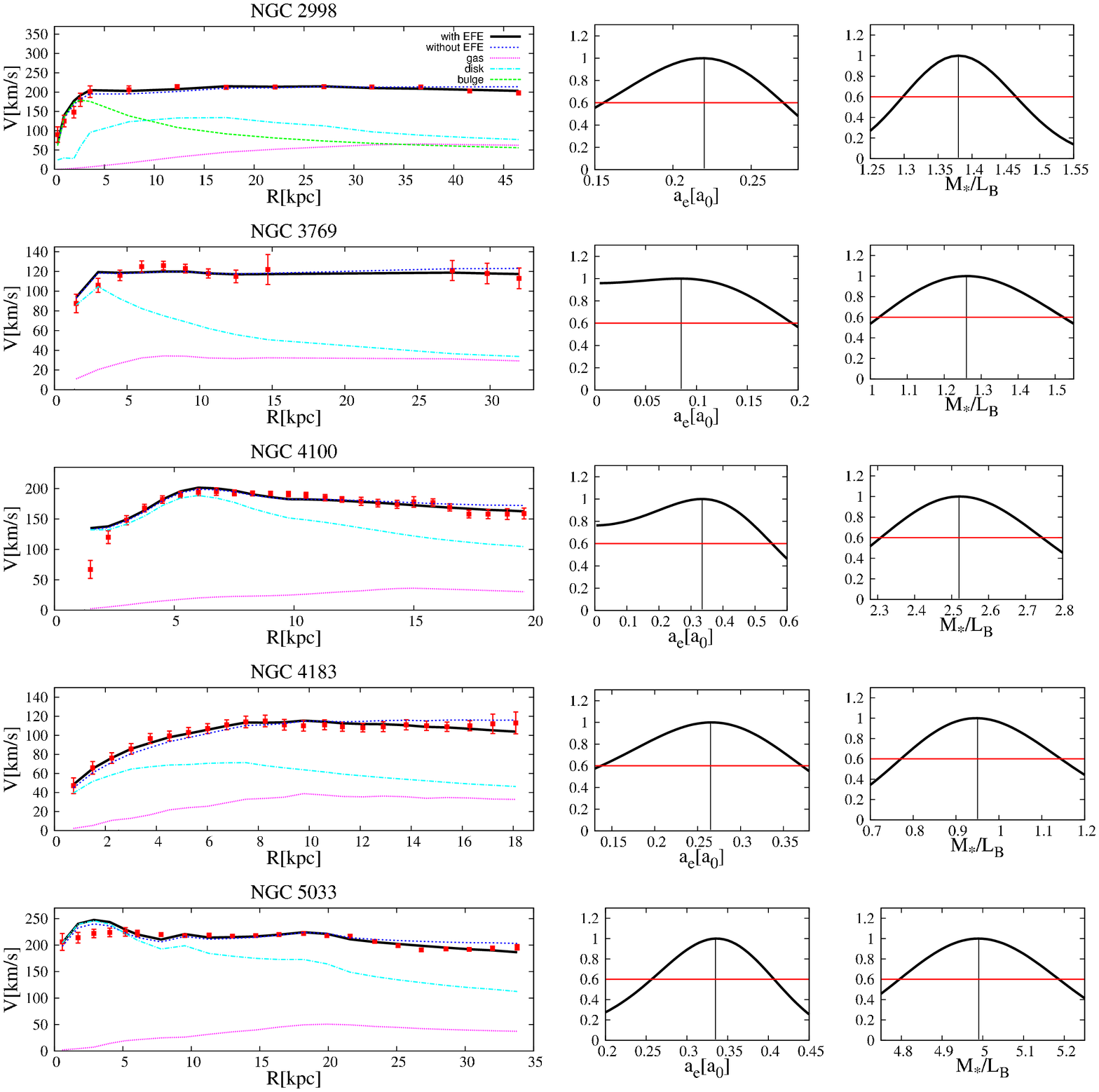}
\caption{Continued.
} \label{f4}
\end{figure*}

\begin{figure*}
\ContinuedFloat
\includegraphics[width=180mm]{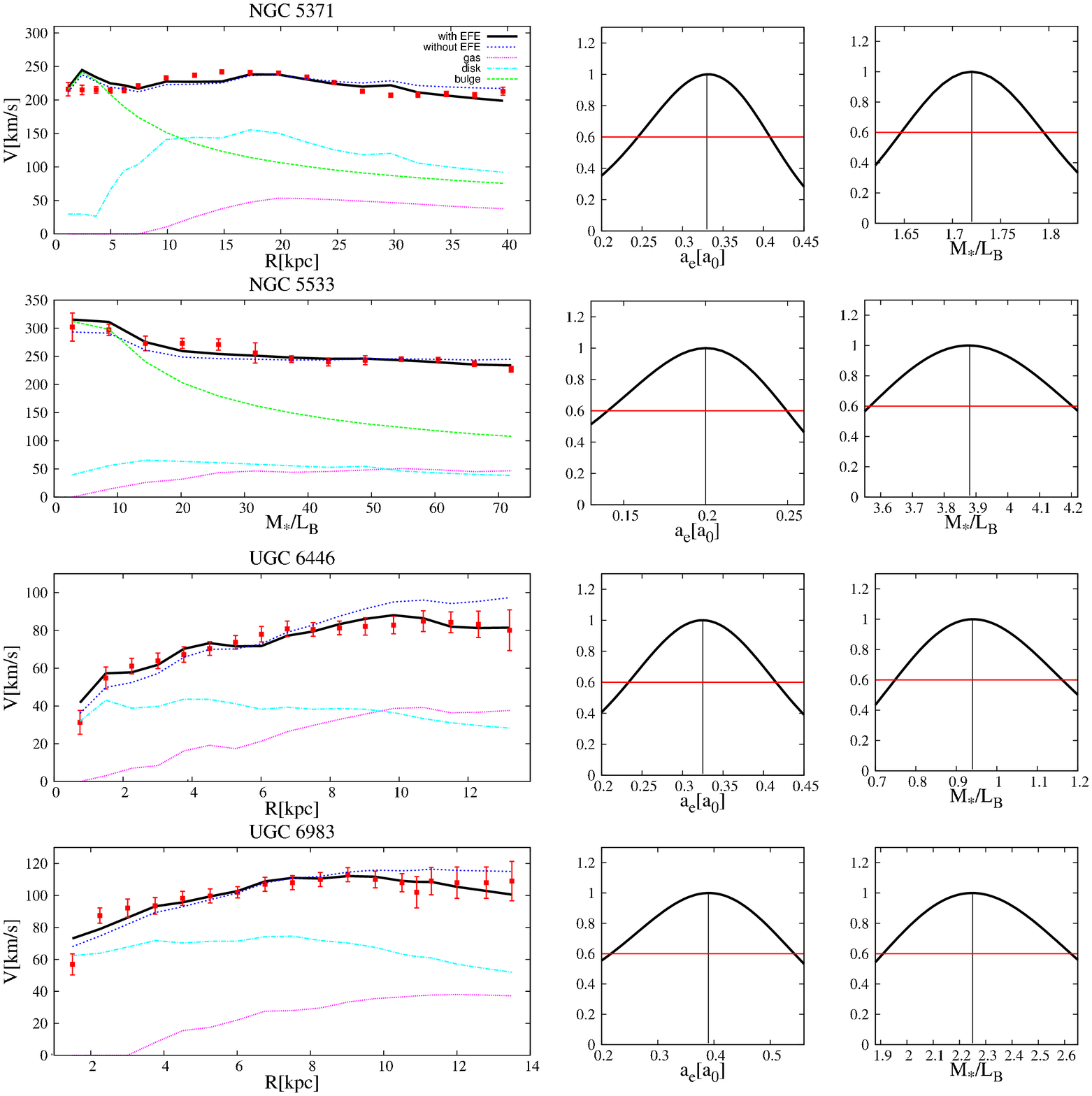}
\caption{Continued.
} \label{f5}
\end{figure*}

\begin{figure*}
\includegraphics[width=180mm]{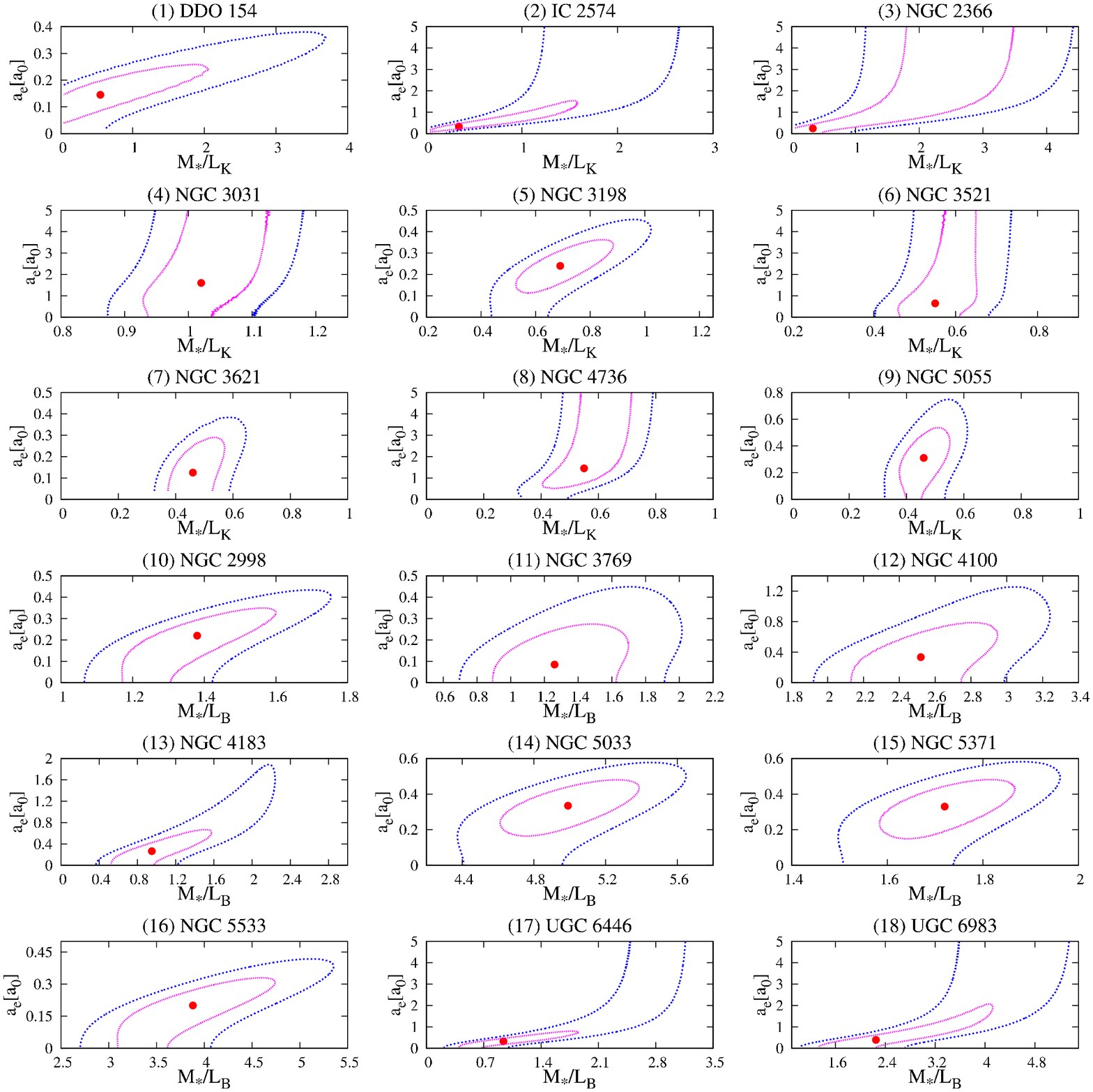}
\caption{ Shown are the  contours of 1 and  2  $\sigma$ joint confidence intervals for the two free parameters: the external field ($a_e$) and the stellar mass-to-light ratio.  The central points indicate the best-fitting values. The details of best-fitting parameters are listed in Tables 1 and 2.
} \label{sigma}
\end{figure*}

In the previous section we found the best-fitting values of the external acceleration, $a_e$, applied to each
individual galaxy. In this section we investigate whether a source of gravitational field can be found in the environment of these galaxies.
However, this is a very difficult task owing to the lack of accurate data on the distances and the masses of perturbing objects. Bellow we discuss the possible sources (perturbers) picked up from the literature for such a gravitational field by searching the environment of individual galaxies in our sample.\\

\textbf{(1) DDO 154}: DDO 154 is a gas-dominated nearby dwarf galaxy with right ascension and declination of $12^h54^m05.2^s$ and $27^d08^m59^s$, respectively. It is located at 3.84 Mpc from us (with the maximum and minimum distance of 4.3 and 3.02 Mpc, respectively \citep{ddo-dm, ddo-md}). The luminosity of the galaxy is about $L_V=4.40 (\pm 0.92)\times 10^7 L_{\odot}$ \citep{ddo-lu}. The measured very low-internal acceleration is in the deep-MOND regime (i.e., $a_{internal}\ll a_0$) and it is free of various uncertainties that occur at higher accelerations. Due to the overestimation of rotational velocities in the outer regions of DDO 154, Milgromian dynamics without the EFE fails to describe the observed rotation curve. Recently, Angus et al. (2012) used a new Milgromian Poisson solver to fit galaxy rotation curves setting the thickness of the gaseous and stellar disk, and also the distance to the galaxy as free parameters and obtained an acceptable Milgromian dynamics fit for DDO 154. Also, in the study by \cite{carignan14}, by letting the Milgromian dynamics standard acceleration ($a_0$) to be free to vary, a good MOND fit is found for a smaller value of $0.68~a_0$. Here, we showed that an external acceleration of about $a_e=0.145^{+0.035}_{-0.030} a_0$ significantly improves the quality of the fit and would help to make the Milgromian dynamics prediction more consistent with observation. According to Table 1, the $\chi^2$ value reduces by an order of magnitude. Nearby NGC 4395 \citep{n4395-dm} is the most conspicuous source for providing the external field in this case. With right ascension and declination of ($12^h25^m48.8^s, 33^d42^m49^s$), angular distance of $7.2535^o$ \citep{n4395-p}, and luminosity of $L_V=1.12 \times 10^{9} L_{\odot}$ \citep{ic-lu} , NGC 4395 is placed at about 812 kpc from DDO 154, with maximum and minimum distances of 3.312 Mpc and 382.32 kpc, respectively. Using the Milgromian dynamics formalism we found NGC 4395 can produce an external field of $a_e=2.87\times 10^{-3}a_0$ with maximum and minimum values of $a_e=6.16\times 10^{-3}a_0$ and $a_e=7.1\times 10^{-4}a_0$, respectively. This amount of external field does not match with the obtained value by the best-fitted model. Taking the acceleration from large scale structure of about $0.01 a_0$ \citep{wu07, wu08} into account, however,  can reduce the inconsistency.

\textbf{(2) IC 2574}: With luminosity about $L_V=8.45 \times 10^{8} L_{\odot}$ \citep{ic-lu} and position of ($10^h28^m23.5^s, 68^d24^m44^s$) in the sky, IC 2574 is located at  the distance of about 3.56 Mpc from us with the minimum and maximum distance estimation of 2.25 and 5.37 Mpc, respectively \citep{ic-dm, ic-md}. The Milgromian dynamics curve predicts significantly higher velocities in the outer regions of this galaxy. Taking the EFE into account, our results show a good improvement in the Milgromian dynamics fits such that the reduced $\chi^2$  decreases from about 3.9 to 1.1. The best fit is obtained with an external acceleration of $a_e=0.330^{+0.090}_{-0.075} a_0$. NGC 3031 is the most important source that we find near IC 2574. With a luminosity of $L_V=2.04\times 10^{10}L_\odot$ \citep{n3031-lu}, right ascension of $9^h55^m33.2^s$ and declination of $69^d03^m55^s$ \citep{n3031-p} the mean distance to NGC 3031 is about 3.71 Mpc from us (with the minimum and maximum distance estimation of 1.4 and 5 Mpc \citep{n3031-dm, n3031-md}, respectively). The angular distance between two galaxies is $4.055^o$ which corresponds to 292.8 kpc radial separation distance  (with the minimum and maximum distance estimation being 159.2 kpc and 3.97 Mpc, respectively). Therefore based on MOND, there is an external field of $a_e=3.43\times 10^{-2}a_0$ (with maximum and minimum values of $a_e=6.31\times 10^{-2}a_0$ and $a_e=2.53\times 10^{-3}a_0$, respectively) at the location of IC 2574 which does not match with the required  external field for this galaxy.

\textbf{(3) NGC 2366}: With the reduced $\chi^2$ value of 0.54, the Milgromian dynamics curve without the EFE is in good agreement with the observed rotation curve of this dwarf galaxy, except that in the middle parts Milgromian dynamics predicts somewhat higher rotational velocities. The measured distance to NGC 2366 is 3.34 Mpc with the minimum and maximum estimation of 1.75 and 6.15 Mpc \citep{ic-dm, ic-md} and the spatial location of this galaxy is determined by right ascention $7^h28^m54.6^s$ and declination $69^d12^m57^s$. As can be seen in Table 2, adding the external field effect slightly improves the MOND fit but significantly increases the best-fitted $M_*/L_K$ ratio from 0.04 to 0.33 in solar units. This new $M_*/L_K$ ratio is more consistent with the SPS prediction than the extremely low value inferred with MOND without the EFE.  The inferred external field from the rotation curve fit is $a_e=0.240^{+0.125}_{-0.155} ~a_0$. The nearest object to this galaxy is NGC 2363 \citep{n2363, n2363-lu} at the angular distance of $0.0182^o$. According to the large uncertainty in the distance of the two galaxies from us the external acceleration induced by NGC 2363 at the location of NGC 2366 covers a large range between $a_e=2.02\times 10^{-5}~a_0$ and $0.108~a_0$ with the central value of $a_e=8.8\times 10^{-5}~a_0$ such that it has an overlap with the external field obtained from the rotation curve.

\textbf{(4) NGC 3031}: The MOND curve predicts significantly higher velocities in the outer regions of this galaxy. Even including the EFE, NGC 3031  still exhibits a large discrepancy between the rotation curve predicted by MOND and the observed rotation curve. This is because of an inner bump in the observed data (at about $R=7.5$ kpc) of this galaxy which is not possible to reproduce in MOND. This could be accompanied by non-circular motions. As given in Table 2, the reduced $\chi^2$ is slightly decreased from 4.3 to 3.3 when the EFE is taken as a free parameter. The large required EFE, $a_e=1.600^{+2.300}_{-0.965} ~a_0$, however makes it hard to find a suitable source for it. NGC 3034 \citep{n3034-lu} is the most considerable neighbor which is located at right ascension $9^h55^m52.7^s$ and declination $69^d40^m46^s$ \citep{n3034-p} in the sky with a distance of about 4 Mpc from us (with a minimum and maximum estimation of 3.2 and 5.5 Mpc \citep{n3034-dm, n3034-md}). There is a $0.3692^o$ angular distance between the two galaxies and we can estimate 308 kpc distance with a maximum value of 4.1 Mpc and a minimum value of 20.6 kpc, between them. The produced external field by NGC 3034 at the location of NGC 3031 is  about $2.14\times 10^{-2} ~a_0$  with the maximum and minimum values of $0.33~a_0$ and $1.61\times 10^{-3}~a_0$, respectively. Although this does not have an overlap within the $68\%$ confidence interval, but this amount of external field covers the $95\%$ confidence interval.

\textbf{(5) NGC 3198}: This galaxy with a Cepheid distance of about 14 Mpc \citep{n3198-dm, n3198-md} has been studid extensively in the context of Milgromian dynamics \citep{carignan14}. Many authors have shown a poor MOND fit for this galaxy, however, adopting the  very unlikely smaller distance of 8.6 Mpc, or by adopting a lower value than the standard one for $a_0$,  one can reconcile MOND with the observed rotation curve \citep{bot02, gen11}.  A good MOND fit can also be obtained by letting the distance vary within the uncertainties \citep{gen13}.  As shown in Figure 2 and Table 1, MOND without the EFE  has a poor quality fit for NGC 3198 with a $\chi^2$ value of 5.49. This is because the MOND curve predicts a higher rotational velocity in the outer parts, which implies more mass, while it underestimates it in the inner part (between 3 and 15 kpc). Applying the EFE, our results show an excellent improvement in MOND fits by reducing the $\chi^2$ value to 1.3, with a required external field of about $a_e=0.24\pm0.05 ~a_0$. This amount of external gravitational field is compatible with the possible external source that we have found in the literature. The nearest object to NGC 3198 is \emph{SDSS J101950.83+453208.0} with luminosity of $L_V=1.3 (\pm0.25) \times 10^{9} L_{\odot}$ and with a distance of about 13.7 Mpc from us  at an angular distance of $0.0066^o$ \citep{ n3034-lu, n4395-p, Merch05}. Such a small angular distance leads to a very large external field estimation induced on NGC 3198  by the galaxy sourcing the external field.  This galaxy can produce an external filed of $a_e= 8.14\times 10^{-3} ~a_0$  with the maximum and minimum values of $a_e= 3.96 ~a_0$ and $5.24\times 10^{-4} ~a_0$, respectively, at the location of NGC 3198.

\textbf{(6) NGC 3521}: Given no significant change in the reduced $\chi^2$, the disagreement between the MOND rotation curve and the observed one remained even when the external field is taken as a free parameter in the fitting process. But as is shown in Figure 2 and Table 1 the existence of an external field with a magnitude of about $0.645^{+0.760}_{-0.645}~a_0$ gives us a better RC compared to the fit without an EFE. \emph{SDSS J110440.08+000329.6} with luminosity of $L_V=2.8 (\pm0.58) \times 10^{7} L_{\odot}$ \citep{n3521-jj} at an angular distance of $0.1707^0$ and a corresponding radial distance of about 1.2 Mpc (maximum 9.8 Mpc and minimum 21.5 kpc) to NGC 3521 \citep{ n3031-dm, n3521-p, n3521-lu, n3521-md} is a good candidate to be the source of the external field. This galaxy can produce an external field of $a_e=3.37\times 10^{-3} a_0$ with a maximum and minimum value of $a_e=1.82\times 10^{-2} a_0$  and $a_e=4.12\times 10^{-5} a_0$ at the location of NGC 3521.

\textbf{(7) NGC 3621}: The MOND rotation curve without the EFE agrees with the observed curve, but an external field of about $a_e=0.125^{+0.090}_{-0.125} ~a_0$ can improve the quality of the fit such that the $\chi^2$ decreases from 1.7 to 1.5. $ESO 377-G 003$ with a luminosity of $L_V=3.9 \times 10^{8} L_{\odot}$ is the nearest gravitational source to NGC 3621 and is located at ($11^h03^m55.2^s, -34^d21^m30^s$).  With an angular distance of 2.52 degree, $ESO 377-G 003$ can produce a gravitational acceleration of about $a_e=5.9 \times 10^{-4} a_0$ at the location of NGC 3621. Due to the large uncertainty in the distance of NGC 3621, the minimum and maximum distance to $ESO 377-G 003$ are 386 kpc and 4.8 Mpc, respectively \citep{n3521-jj, n3198-dm} which leads to a gravitational acceleration between $a_e=4.0\times 10^{-3} a_0$ and $3.2 \times 10^{-4} a_0$. Therefore, the gravitational acceleration imposed by the nearby source has an overlap with the external field obtained from the  MOND fit rotation curve.


\textbf{(8) NGC 4736}: The Milgromian dynamics curve without the EFE agrees well with the observed curve in the inner parts, but it predicts significantly higher-than-observed velocities in the outer parts. The EFE leads to a velocity decrease in the external part of the rotation curve and significantly improves the quality of fit such that the chi-squared reduces from  $\chi^2=5.96$ to $1.57$. A best fitted external filed  of $a_e=1.450^{+1.350}_{-0.505} ~a_0$  is required for this agreement between MOND and the observation. This galaxy is located at a distance of about $5.02^{+1.4}_{-0.8}$ Mpc from us.  The galaxy NGC 4244 \citep{ic-md, n4244-dm} is the most effective source near to NGC 4736 \citep{n3034-dm, n4736-p, n3034-lu, n4736-dm} with a distance of 4.12 Mpc from us. With a luminosity of $L_V=2.82\times 10^9L_{\odot}$ and distance of $0.98^{+5.18}_{-0.59}$ Mpc  to NGC 4736, it can produce an external field of $a_e=3.81^{+5.86}_{-3.20}\times10^{-3} ~a_0$. This is far from the predicted external field. Note that the discrepancy can be reduced if the external gravitational field of about $a_e=10^{-2} a_0$ exerted by the large scale structures being considered.

\textbf{(9) NGC 5055}: Although the general form of the MOND rotation curve agrees with the observed curve, the latter is not reproduced well in the outer parts where the predicted rotation curve by MOND without an EFE falls above the observed one. An external filed of about $a_e=0.310^{+0.115}_{-0.135} a_0$ can improve the quality of the fit such that the $\chi^2$ decreases from 3.06 to 1.15. NGC 5194  \citep{n5194-p, n3034-md, n3034-lu}  within $5.315^o$ angular distance from NGC 5055 which corresponds to the radial distance of about 855 kpc (maximum 7.215 Mpc and minimum 463 kpc) can produce $a_e=1.2\times10^{-2}~a_0$ (maximum $a_e=2.22\times10^{-2}a_0$ and minimum $a_e=1.43\times10^{-3}a_0$) at the location of NGC 5055 \citep{n3198-dm, n5055-lu, ic-md} which has a good overlap with the implied $a_e$ by the Milgromian dynamics fit within the $95\%$ confidence interval.

\textbf{(10) NGC 2998}: This galaxy with right ascension and declination of ($9^h48^m43.6^s, 44^d04^m53^s$) is located at a distance of about $61^{+8}_{-16}$ Mpc from us , and the luminosity of the galaxy is $L_V=2.69\times 10^{10} L_\odot $ \citep{n2998-dm, n3521-lu, n2998-md}. A better RC fit with an external field of $a_e=0.220^{+0.050}_{-0.065} ~a_0$ is achieved. This external field can reduce $\chi^2$ from 2.7 to 1.2. NGC 3008 is probably a good candidate as a source of the external field for NGC 2998. NGC 3008 \citep{n3521-jj, n3008-jj} with a luminosity of $L_V=1.33 (\pm 0.27)\times 10^{10}$ produces an external field of about $a_e=1.06\times 10^{-3}a_0$ (with the maximum and minimum values of $a_e=8.2\times 10^{-2}a_0$ and $a_e=2.83\times 10^{-4}a_0$, respectively), which has a good  overlap with theoretical value within the $95\%$ confidence interval.

\textbf{(11) NGC 3769}: This galaxy with the total luminosity of $L_V=4.7 \times 10^9 L_{\odot}$ \citep{n3521-lu} and  right ascension  and declination of ($11^h37^m44.1^s, 47^d53^m35^s$) is located at the distance of $16^{+4}_{-6}$ Mpc \citep{n2998-md, n3769-dm}. Figure 2 shows that the observational RC falls slowly at large distances but that the MOND RC without an external field is almost constant. A sufficient fall in the RC is achieved by adding an external field equal to $a_e=0.085 a_0$ with the maximum and minimum values of $a_e=0.195 a_0$ and zero, respectively.  NGC 3769A \citep{n3769a-lu} is the most considerable nearby galaxy to NGC 3769 which can produce enough gravitational acceleration at the location of NGC 3769. The mean distance of NGC 3769A from us is 17 Mpc \citep{n3031-dm} with a projected distance of about $0.0108^o$ to NGC 3769, which is roughly enough to produce an external gravitational field of  $a_e=1.53 \times10^{-3} a_0$ with the maximum and minimum values of $a_e=0.472a_0$ and $a_e=2.06 \times 10^{-4} a_0$ in the  MOND regime which is compatible with the value obtained from the MOND fit with an EFE. We assume that the  Mass to Light ratio of NGC 3769A in the V-band is 4 as an upper limit.

\textbf{(12) NGC 4100}: The luminosity of this galaxy is about $L_B=2.12 \pm 0.90 \times 10^{10} L_{\odot}$, and it is located at right ascension  and declination of ($12^h06^m08.4^s, 49^d34^m58^s$) \citep{n4100-lu}. The measured distance to NGC 4100 is about  21.8 Mpc with the minimum and maximum estimation of 15.7 Mpc and 25.5 Mpc \citep{ic-md, n3521-md}, respectively. As can be seen in Figure 2 the observed rotation curve of this galaxy falls  bellow  the curve predicted by MOND without an EFE in the outer part.   Adding an external field of $a_e=0.335 ~a_0$  improves the fit by reducing the $\chi^2$ value from 2.15 to 1.87. The nearest object to NGC 4100 is  \emph{SDSS J120608.59+493459.9} which is located at ($12^h06^m8.6^s, 49^d35^m00^s$) with a distance of 17.8 Mpc from us \citep{2004sdss}. Therefore with an angular distance of about $0.0042^o$ to NGC 4100 this object produces an external gravitational field of about $a_e=1.08 \times 10^{-4}  ~a_0$ (with maximum $a_e=0.285 ~a_0$ and minimum $a_e=3.45 \times 10^{-5}  ~a_0$), in the MOND regime, at the location of NGC 4100. This amount of external field is in agreement with the external field that obtained from the MOND fit within the $68\%$  confidence interval.

\textbf{(13) NGC 4183}: The overall form of the MOND RC without an EFE  agrees well with the observed data with a reduced $\chi^2$ value of 1.01.  The rotation curve fit is improved by the addition of an EFE (with the best-fitted value of  $a_e=0.265^{+0.110}_{-0.130} a_0$ for the external filed) and leads to a lower $\chi^2$ value  of 0.27. NGC 4138 \citep{n4138-md, n4138-dm, n3521-lu} is the most important galaxy which can gravitationally affect NGC 4183 \citep{n2998-dm, n2998-md}. With a luminosity of $L_V=1.93\times 10^{10}L_\odot$ and angular distance of $0.4217^o$ to NGC 4183, this galaxy produces an external field of $3.29\times 10^{-3}~a_0$ (with the maximum and minim values of $9.64\times 10^{-2}~a_0$ and $4.27\times 10^{-4}~a_0$, respectively) which is compatible with theoretical prediction within the $95\%$ confidence interval.

\textbf{(14) NGC 5033}: With a luminosity of about $L_V=3.91 \times 10^{10} L_{\odot}$ \citep{n3521-lu}, NGC 5033 is located at ($13^h13^m27.4^s, 36^d35^m38^s$ \citep{n4395-p}) and at a distance of about 19.6 Mpc from us with a large uncertainty such that the minimum and maximum distance estimation is 15.2 and 42.5 Mpc \citep{n3198-dm, n5033-md}, respectively. This is one of the galaxies that MOND could not reproduce acceptably with a reduced $\chi^2$ of 7.04. MOND overestimates the rotation velocities in the inner parts and underestimates them in the outer parts. A better fit with $\chi^2=3.1$ is obtained by including the EFE. The required external field is $a_e=0.335^{+0.070}_{-0.080} a_0$. However, Figure \ref{f2} shows that even including the EFE, NGC 5033 still exhibits a large discrepancy between the MOND RC and the observed RC in the central region. NGC 5014 with a luminosity of $L_V=(2.30 \pm 0.67) \time 10^{11} L_{\odot}$, located at ($13^h11^m31.2^s, 36^d16^m56^s$) with a distance of $19.26^{+3.54}_{-1.86}$ Mpc from us is the nearest known gravitational source to NGC 5033.  With an angular distance of $0.3027^o$, NGC 5014 can produce a gravitational acceleration of about $a_e=9.72 \times 10^{-2} a_0$ at the location of NGC 5033. Due to the large uncertainty in the distance of NGC 5033, the minimum and maximum distances to NGC 5014 are 90 kpc and 25.5 Mpc, respectively, which leads to a gravitational acceleration between $a_e=0.436a_0$ and $1.5 \times 10^{-3} a_0$. Therefore, the implied gravitational acceleration from the MOND fit is in agreement within the error bars with the gravitational acceleration imposed by the nearby source.

\textbf{(15) NGC 5371}: This is one of the galaxies for which  MOND could not produce an acceptable fit to the observed rotation curve. MOND overestimates the rotation velocities in the inner parts of the rotation curve which leads to a high reduced chi-squared of $\chi^2=10.49$.  A slightly better fit is achieved by introducing the external field effect into the fit. The best-fit is obtained by $a_e= 0.330^{+0.080}_{-0.085} $. With a luminosity of $L_V=3.73\times 10^{10}L_\odot$, NGC 5353 \citep{n5353-1} lies $0.267^o$ from NGC 5371 and with a radial distance of 4.3 Mpc (maximum 14.8 Mpc and minimum 129.5 kpc) it can produce an external gravitational field of $a_e=3.17\times 10^{-3}a_0$, with maximum and minimum values $a_e=0.105a_0$ and $a_e=9.18\times 10^{-4}a_0$ respectively. However the minimum required gravitational acceleration (i.e., $0.245 a_0$ )  is not consistent with the maximum external field (i.e., $a_e=0.105 a_0$) at the 68\% confidence level, but according to Fig. 6, it is marginally compatible  within the $95\%$ confidence interval. Assuming 10\% uncertainties in the distance of this galaxy and using the minimum limit for the distance \citep{Cap11}, the best-fit is obtained by $a_e= 0.300^{+0.175}_{-0.290} $ which is in good agreement with the imposed gravitational acceleration by the nearby source within the 68\% confidence interval.

\textbf{(16) NGC 5533}: This galaxy with a distance of $51.22^{+5.98}_{-11.02}$ Mpc \citep{n3521-lu, n2998-md, n3521-md} from us is located at the celestial coordinates of ($14^h16^m07.7^s, 35^d20^m38^s$). Applying the EFE, our results show a good improvement in Milgromian dynamics fits such that the reduced $\chi^2$  decreases from about 2.5 to 1.1. The best fit is obtained with an external acceleration of $a_e=0.20^{+0.05}_{-0.06}$.  \emph{SDSS J141606.29+352022.5} \citep{1-n5533} is a galaxy that can induce the necessary external field. The celestial coordinates of this system are ($14^h16^m06.3^s, 35^d20^m23^s$) and it is located at 54.0 Mpc from us with 10\% uncertainty. According to the projected distance of $0.002^o$ to NGC 5533 the minimum and maximum estimated distances between the two galaxies are 1.64 kpc and 19.2 Mpc, respectively, with a mean value of  2.78 Mpc. Therefore, with a luminosity of $L_V=2.43 \times 10^{9} L_{\odot}$ it can produce $a_e= 1.25\times 10^{-3} a_0$ with the maximum and minimum values of $a_e= 4.55 a_0$ and $1.81\times 10^{-4} a_0$, respectively, which is in good agreement with the implied external field by the Milgromian dynamics fit.

\textbf{(17) UGC 6446}: The rotation curve as predicted by Milgromian dynamics without the EFE falls bellow the observed one in the inner parts of the rotation curve (except the first data points in the central region), and above the observed rotation curve in the outer parts. So, an external field would help to make the Milgromian dynamics prediction more compatible with the observed curve.  Milgromian dynamics without the EFE reproduces the observed RC poorly (with a reduced $\chi^2$ of about 2.43), while the Milgromian dynamics curve with an EFE is in excellent agreement with the observed rotation velocities. A good MOND fit is found for a value of $a_e= 0.325^{+0.090}_{-0.090} a_0$.  This galaxy is on the edge of the boundary taken to define the UMa cluster in both velocity and angle on the sky. In fact the Ursa Major Cluster is the poorest known galaxy cluster, with a velocity dispersion of only 148 $kms^{-1}$ and a virial radius of 880 kpc \citep{n3769a-lu}. The galaxies are distributed with no particular concentration toward any center. We assumed the Ursa Major Cluster to be a sphere with a luminosity of $L_V=4.16 \times 10^{11} L_{\odot,V}$  and involving gas assuming the virial theorem given the observed velocity dispersion \citep{n3769a-lu}. Therefore, using equation 1 we found that the cluster, in the MOND regime, can produce an external gravitational field of $a_e= 5.58\times 10^{-2}a_0$ at the half radius and $a_e= 7.89\times 10^{-2}a_0$ at the edge of the sphere. As can be seen in Fig. 7 the value of $a_e$ imposed by the host galaxy group is marginally in agreement with the uncertainty with the value found by the Milgromian dynamics fit given in Table 2.  A MOND fit with distance let free to vary within the uncertainties \citep{Tul08} is obtained with an external acceleration of $a_e= 0.325^{+0.210}_{-0.289} a_0$ which matches with the produced external field for this galaxy within the 1$\sigma$ confidence level.

\textbf{(18) UGC 6983}: This galaxy is also a member of the Ursa Major Cluster and has the same behavior in its rotation curve. The Milgromian dynamics fit without the EFE gives us $\chi^2=1.37$ but by using an external field as a free parameter in the rotation curve fit the $\chi^2$ reduces to 0.79 and a better fit is achieved with an external field of $a_e=0.390^{+0.155}_{-0.175}$. The Ursa Major Cluster can produce a strong enough gravitational field as a source of the EFE. As we can see in Figure 2 the amount of external field is in good agreement within the $95\%$ confidence interval with the theoretical value implied by the MOND RC fit.

\begin{figure*}
\includegraphics[width=85mm]{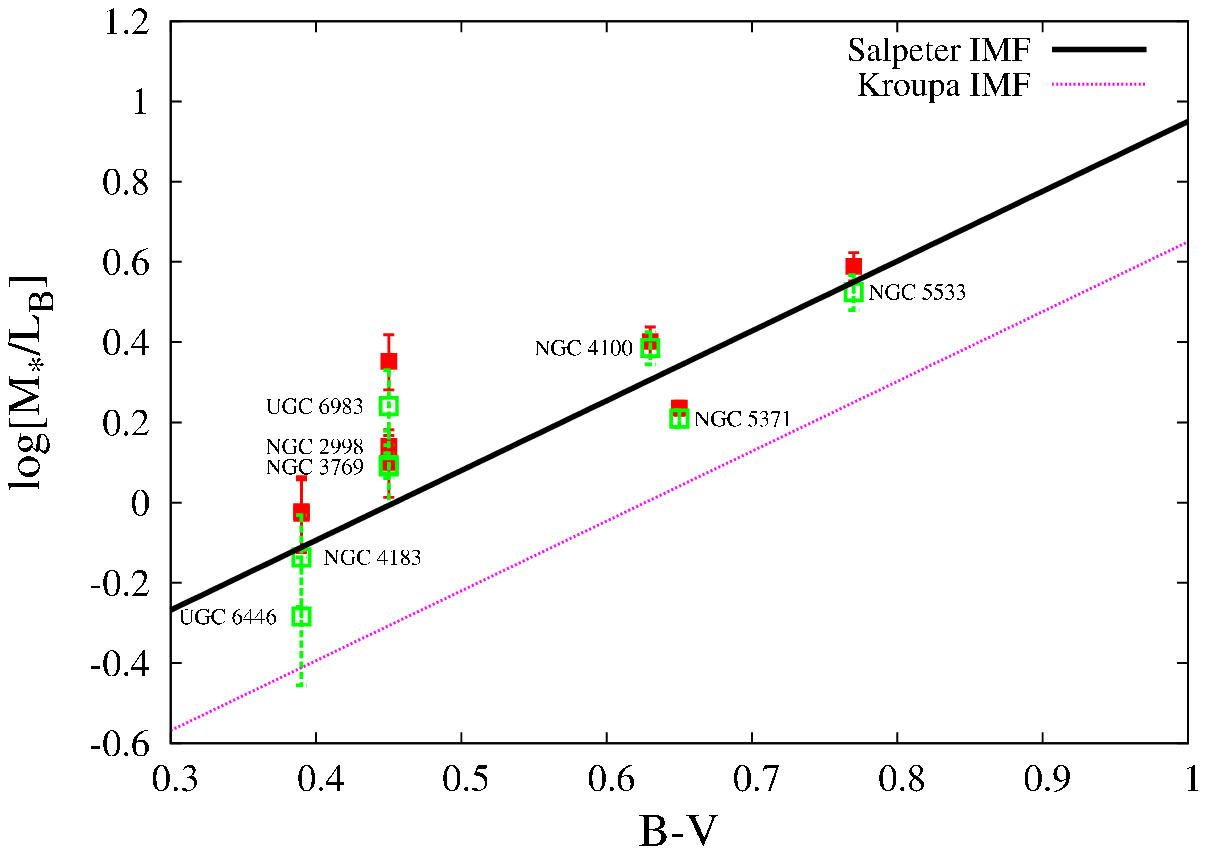}
\includegraphics[width=85mm]{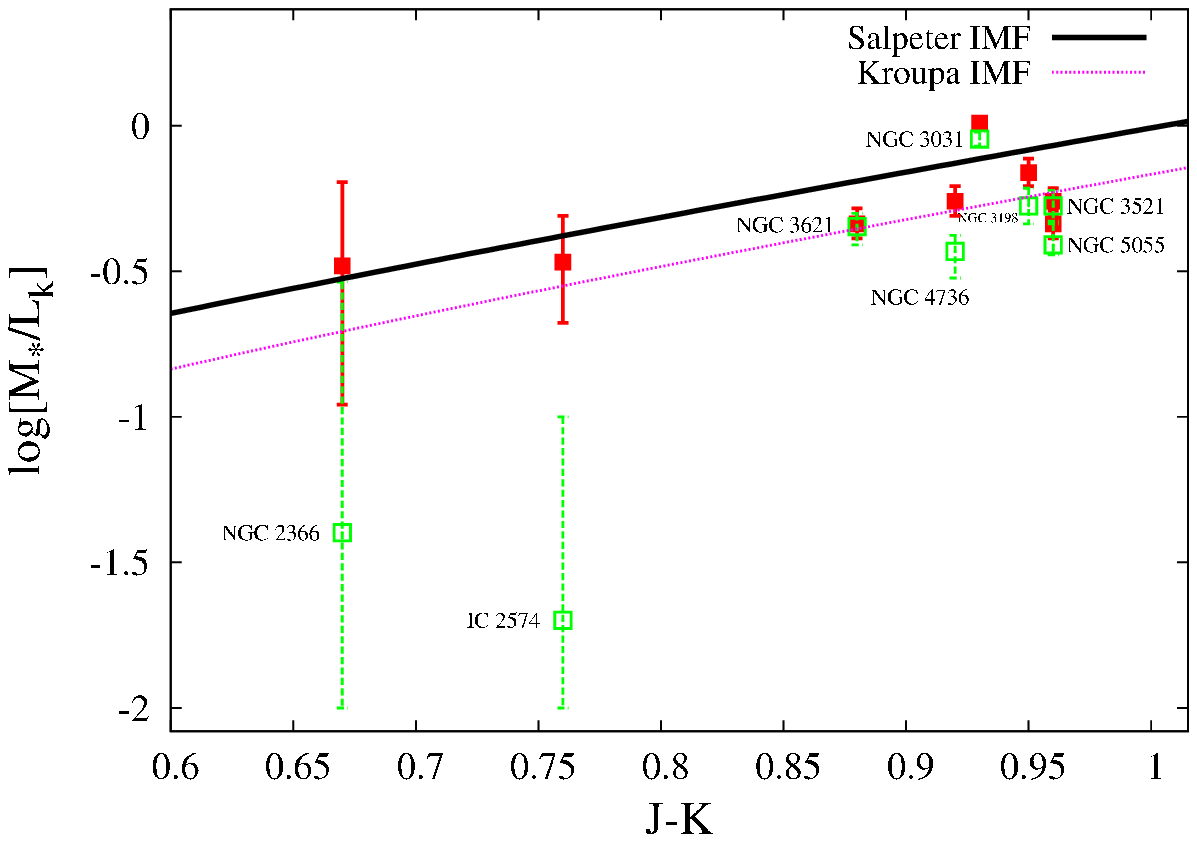}
\caption{A comparison of the best fit stellar $M_*/L$ ratios obtained from MOND rotation curve fits with the independent expectations of stellar population synthesis models (lines). Plots of stellar $M_*/L_B$ ratios versus B-V color (left panel) and $M_*/L_K$ ratios versus J-K color (right panel). The corresponding data points for galaxies NGC 5033 and DDO 154 are not plotted due to the lack of information about the color. The red filled squares represent the implied $M_*/L$ values with including the EFE and the green open squares shows the predictions of MOND fits without the EFE. Solid lines in each panel denote the theoretical predictions of SPS with different IMFs. They have almost the same slope but different y-intercepts. We note that the galaxy-wide IMF, the IGIMF, is predicted to be top-light for the galaxies in the present sample \citep{weid05, weid13} which qualitatively corresponds to a bottom-heavier IMF than the canonical IMF \citep{Kroupa01, kro13} which is likely the reason why the red data points appear to be better represented by a Salpeter IMF.
} \label{f6}
\end{figure*}

\begin{figure}
\includegraphics[width=85mm]{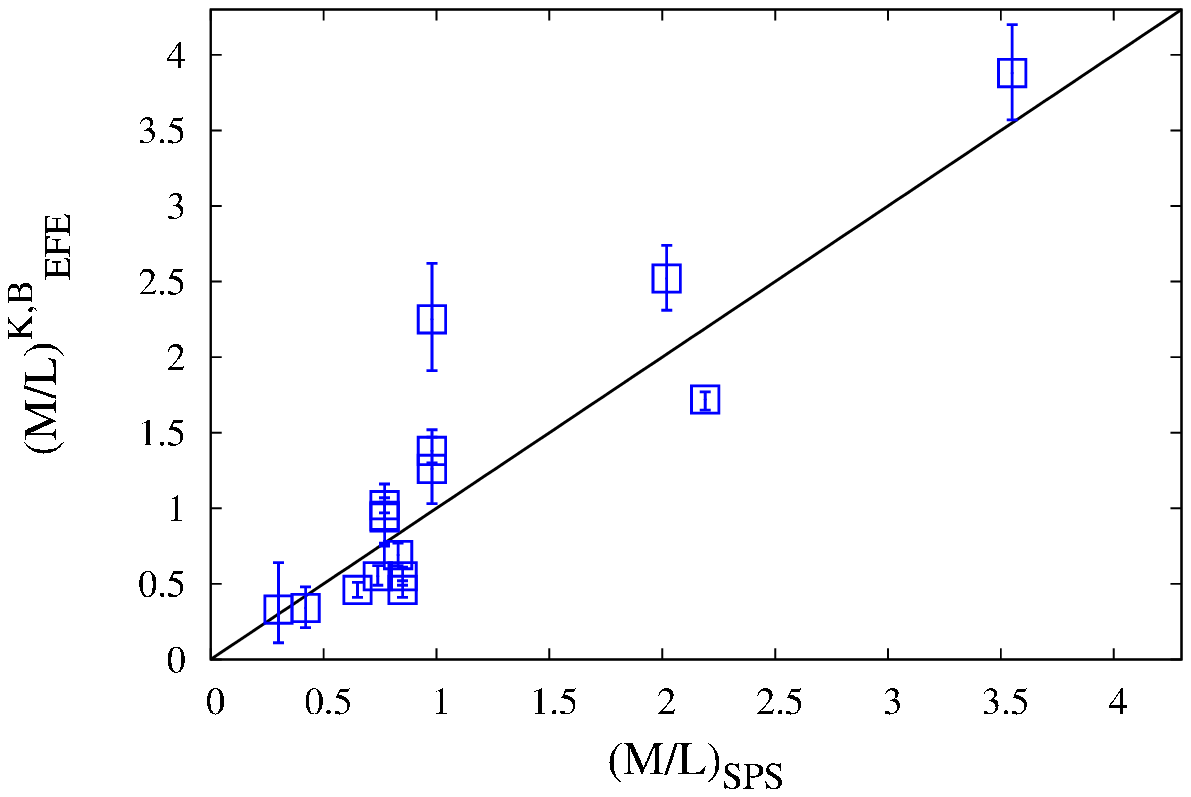}
\caption{The best fit stellar $M_*/L$ ratios obtained from MOND rotation curve fits with EFE vs. SPS prediction with a Salpeter IMF which is a first-order approximation to the top-light IGIMF \citep{weid05, weid13, kro13}. The  data points for galaxies NGC 5033 and DDO 154 are not plotted due to the lack of information about the color.
} \label{sps-efe}
\end{figure}

\begin{figure}
\includegraphics[width=85mm]{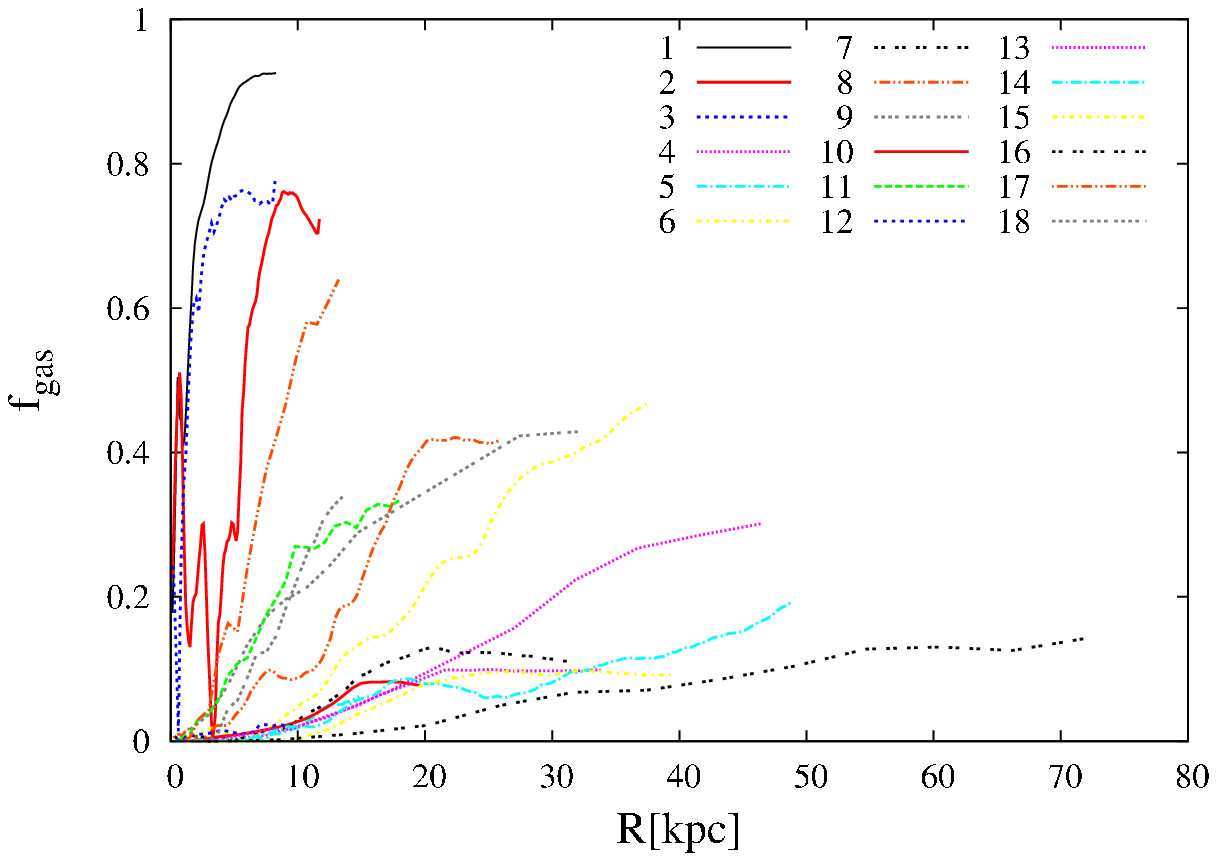}
\caption{Gas fraction within radius R for all galaxies in our sample.
} \label{fgas}
\end{figure}

\subsection{Summary}

In Figure \ref{efe} we compare the implied external acceleration from Milgromian dynamics fits with the EFE and those estimated from the nearby galaxy or cluster of galaxies. As can be seen in 15 out of 18 galaxies the implied accelerations are compatible with what is probably imposed by the external sources within the 95\% confidence level.

Note that we have considered only one possible source for the EFE. In reality each of the galaxies in Table 1 and 2 is immersed in an external field streaming from the entire matter distribution around it such that the $a_e$ values obtained from the one source are likely lower boundaries to the true full $a_e$. One would need to construct a 3D mass distribution of the observed universe to calculate the EFE at every point. This is now possible with the Phantom of Ramses (PoR) code \citep{lug15}.

It is worth mentioning that  there is some inevitable external gravitational field of about $0.01 a_0$ exerted by large scale structure, estimated from the acceleration endured by the Local Group during a Hubble time in order to attain a peculiar velocity of 600 km/s. Although, this acceleration is not enough to remedy the problematic cases such as DDO 154 and NGC 4736 that the imposed $a_e$ by external sources is slightly less than $0.01a_0$, it can reduce the amount of discrepancy for these galaxies.

Figure \ref{md} depicts the distance between the galaxy souring the required external gravitational field  and the galaxies in our sample, $d$, vs. their baryonic masses. The lower, central and upper values of the mass of the perturber are calculated from their intrinsic luminosity assuming $M_*/L=$ 1, 2, and 4, respectively.  For a given $a_e$, one can calculate $d$ as a function of perturber mass, $M_{pert}$ as follows \citep{wu15}:

\begin{equation}
	d = \frac{\sqrt{G M_{pert}a_0}}{a_e},
\end{equation}
where, $M_{pert}$  is the baryonic mass of nearby gravitational source. For a given acceleration $a_e$, in the range $[10^{-3} - 1] a_0$  we plotted the relation  $d(M_{pert})$  as straight lines in Fig. \ref{md}.

\section{Conclusions} \label{conclusion}

It is well accepted that Milgromian dynamics is successful in reproducing observed rotation curves of a large sample of galaxies. But it appear to fail in reproducing the observed rotation curves of some galaxies that show decreasing velocities in their outer parts.  In these galaxies Milgromian dynamics predicts higher velocities in the outer rotation curve than is observed \citep{swaters10}. This systematic deviation can be explained if the Milgromian dynamics acceleration parameter, $a_0$, is slightly lower than usually assumed, but this poses a serious challenge to Milgromian dynamics since  $a_0$ ought to be a universal constant. Milgromian dynamics predicts higher rotational velocities in the inner regions and lower ones in the outer parts when the EFE is taken into account. In this paper we assessed if this incompatibility can be  resolved by taking the second order effect of Milgromian dynamics, i.e., the external field effect of nearby sources, into account.

We presented MOND fits with and without the EFE for a sample of 18 galaxies selected from the literature.  The selected galaxies in the sample cover a large range of luminosities and morphological types. The MOND results including the EFE are:

\begin{itemize}
\item  Taking into account the external gravitational field as a free parameter leads to  a remarkable success in predicting the observed rotation curves in virtually all cases. The quality of the fit is, in about 50\% of  the members in our sample, significantly improved by the reduced chi-squared value being reduced by a factor of 2 to 5. This means that by including the EFE, which is a required physical component for any non-isolated galaxy in Milgromian dynamics, the whole form of the rotation curve changes to naturally reproduce the observed curve.
\item We showed that the EFE can successfully remedy the overestimated rotational velocities of galaxies in Milgromian dynamics.
\item Comparing the fitted global stellar mass-to-light ratios to the predictions of stellar population synthesis models of \cite{Oh08} we showed that the implied values in the MOND framework (with the EFE) are  more reasonable than those implied from MOND fits without the EFE.  Indeed, an impressive result from the MOND + EFE modelling presented here is that the $M_*/L$ values required to fit the observed rotation curves are well consistent with those of SPS models, providing a further consistency test that Milgromian dynamics is indeed the correct effective approach in the ultra-weak field gravitational limit. This issue of stellar populations will be addressed in a forthcoming contribution more closely.
\item We used the EFE to constrain the possible source of the gravitational field surrounding each individual galaxy in our sample.
\end{itemize}

We thus propose that including the EFE can be used to resolve the central surface brightness- MOND acceleration constant correlation that has been discussed by some authors \cite{carignan14, swaters10}, in the sense that the galaxies with higher surface brightness require a higher value of $a_0$ and galaxies with a lower surface brightness tend to have lower $a_0$. This is because, adding the EFE or decreasing  $a_0$ have almost a similar impact, and both lead to the rotation curve of a galaxy bending downwards.

It should be noted that our approach in adding the 1D version of the external field effect is a first order approximation. It is worth to investigate the rotational velocities of each individual galaxy using a 3D solution of the quasi linear formalism of MOND (QUMOND) from a numerical Poisson solver. This is readily possible now with the Phantom of Ramses (PoR) code recently publicly released by \cite{lug15}. We leave this task  for an upcoming paper to be investigated in detail (Haghi et al., in preparation).

Considering this sample of rotation curves, overall we see that in about 16 out of the 18 cases the MOND curve closely agrees with the observed RC (with $\chi^2 \leq 4 $). In some cases (e.g., NGC 3198, NGC 5055, and UGC 6446) the MOND curve with the EFE is a spectacular reproduction of the observed curve. In two other cases, however, MOND  correctly predicts the general behavior of the observed curve but misses details.  Such details may result from non-axisymmetrical systems, some non-circular motions, and spiral density waves.

Moreover, including the distance uncertainties in the case of galaxies in which MOND predicts higher values for the external field  compared to the imposed acceleration by nearby sources can help to reduce the discrepancy. Since larger distances have qualitatively the same effect as the external field effect, we showed that using the lower limit of the distance (within the uncertainties) leads to a lower value of the best-fitted external acceleration in the MOND fits. Finally, we point out that the modelling performed here is done using the approximation via Eq. 6 to the true and hitherto not yet discovered theory of gravitation. This approximation to the three-dimensionally-defined rotational structure of disk galaxies may add an uncertainty of a few percent to the fits obtained here.

\begin{figure}
\includegraphics[width=85mm]{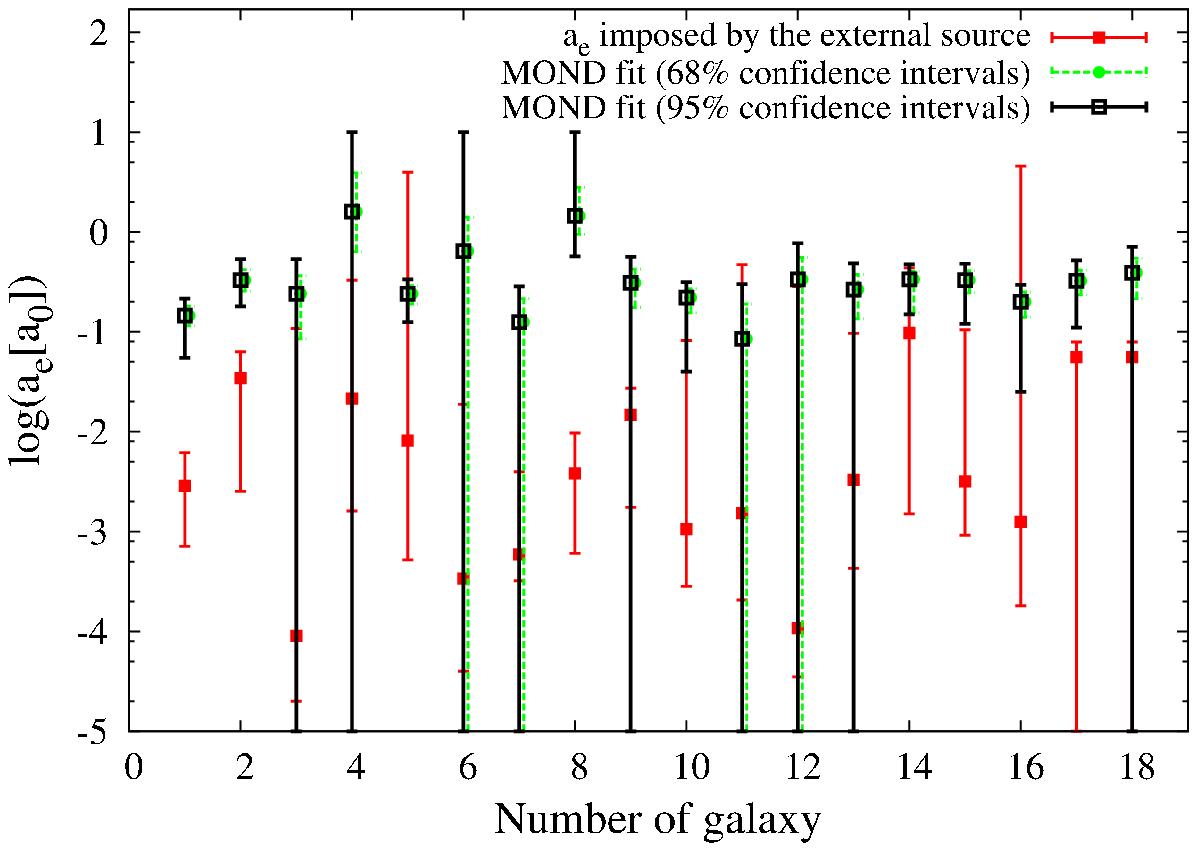}
\caption{ The estimated and predicted external gravitational acceleration of 18 dwarf and spiral  galaxies in the EFE regime. Each number along the x-axis represents the hypothetical ID corresponding to each galaxy (as given in the first column of Tables 1 and 2). Filled red symbols are the estimated external field imposed by the strongest source nearby individual galaxies. The predictions of Milgromian dynamics with the EFE  are shown as black (within 95\%confidence intervals) and green (within 68\%confidence intervals) symbols. Note that in reality there may be other additional sources such that our estimates for $a_e$ are probably lower values only.
} \label{efe}
\end{figure}

\begin{figure}
\includegraphics[width=85mm]{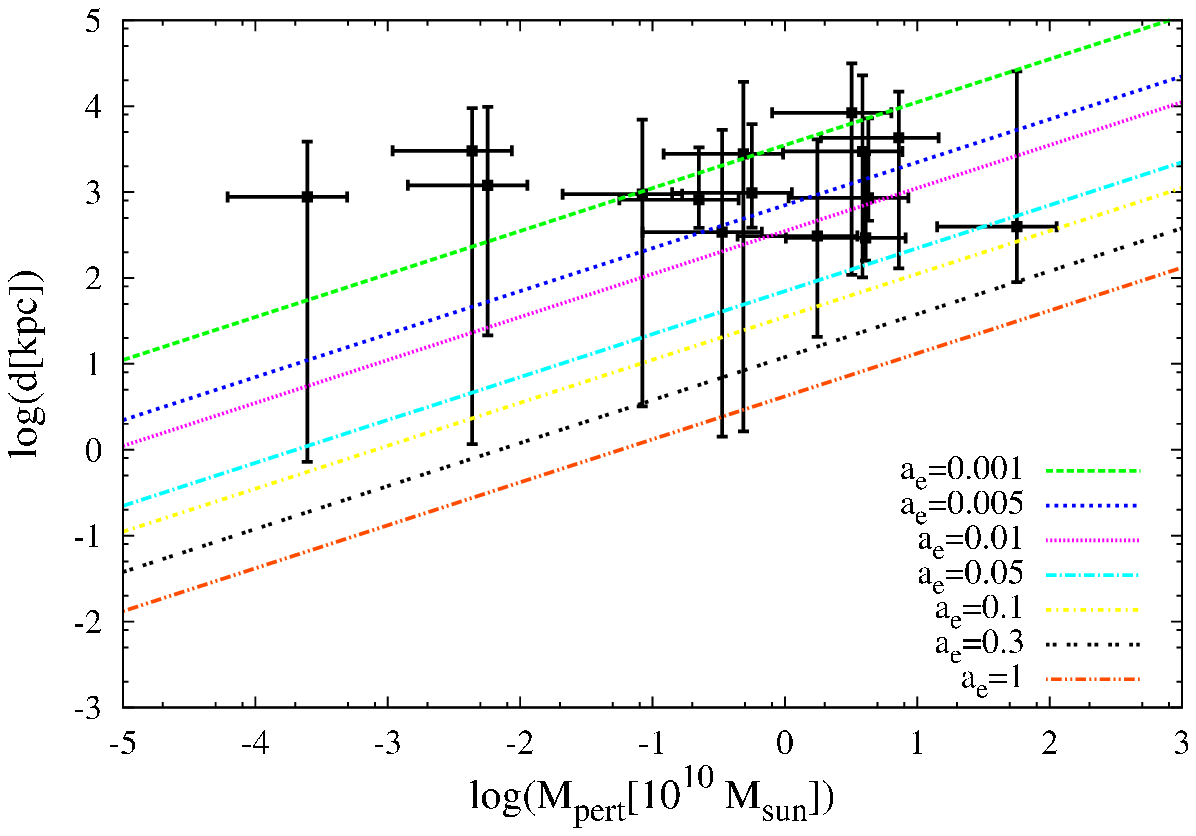}
\caption{ Distance of the galaxy sourcing the gravitational field (external field source) to the center of individual galaxies in our sample versus the baryonic mass of the external field source, which could be a nearby galaxy as given by the black points with uncertainties and discussed in Sec. 6. Different lines represent constant external acceleration contours.
} \label{md}
\end{figure}

%

\bibliographystyle{plainnat}

\bsp \label{lastpage} \end{document}